\documentclass[journal=jpclcd,manuscript=letter]{achemso}
\usepackage[T1]{fontenc}
\usepackage[utf8]{inputenc}
\usepackage{amsmath}
\usepackage{graphicx}
\PassOptionsToPackage{version=3}{mhchem}
\usepackage{mhchem}

\makeatletter

\providecommand{\tabularnewline}{\\}

\providecommand{\tabularnewline}{\\}

\usepackage{commath}
\usepackage{subcaption}
\usepackage{multirow}
\usepackage{booktabs}
\usepackage{silence}
\usepackage{array}
\usepackage{braket}
\usepackage{placeins}
\usepackage{siunitx}
\usepackage{threeparttable}

\author{Jiachen Li}
\affiliation{Department of Chemistry, Duke University, Durham, NC 27708, USA}
\author{Zehua Chen\footnote{Jiachen Li and Zehua Chen share the first authorship.}}
\affiliation{Department of Chemistry, Duke University, Durham, NC 27708, USA}
\author{Weitao Yang}
\affiliation{Department of Chemistry, Duke University, Durham, NC 27708, USA}
\email{weitao.yang@duke.edu}

\title{Multireference Density Functional Theory for Describing
Ground and Excited States with Renormalized Singles}

\makeatother

\begin{document}

\begin{tocentry}
\includegraphics[width=1\textwidth]{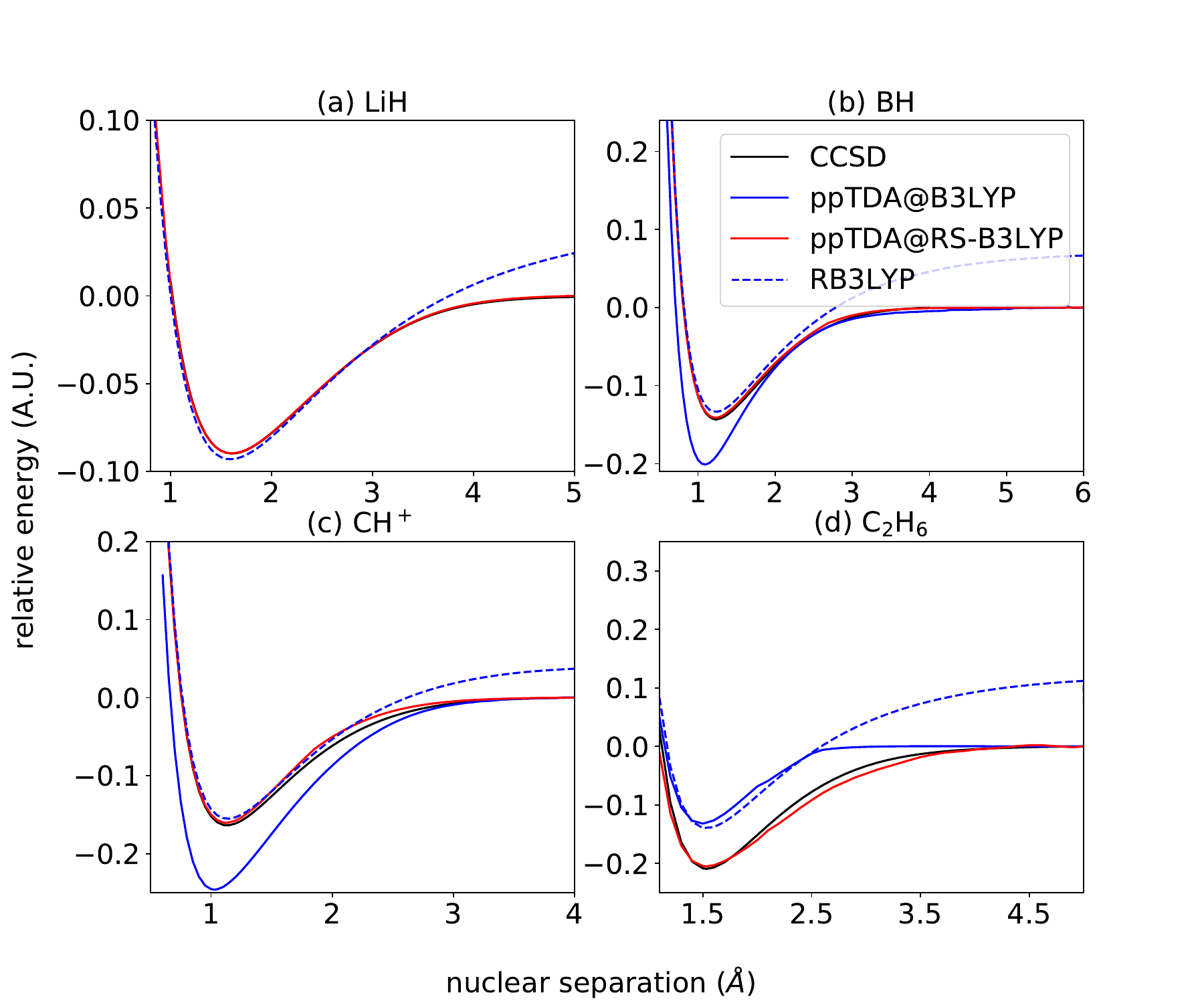}
\end{tocentry}

\begin{abstract}
  We applied renormalized singles (RS) in the multireference density functional theory (DFT) to calculate accurate energies of ground and excited states.
  The multireference DFT approach determines the total energy of the $N$-electron system as the sum of the ($N-2$)-electron energy from a density functional approximation (DFA) and the two-electron addition energies from the particle-particle Tamm-Dancoff approximation (ppTDA),
  naturally including multireference description.
  The ppTDA@RS-DFA approach uses the RS Hamiltonian capturing all singles contributions in calculating two-electron addition energies,
  and its total energy is optimized with the optimized effective potential method.
  It significantly improves the original ppTDA@DFA.
  For ground states,
  ppTDA@RS-DFA properly describes dissociation curves tested and the double bond rotation of ethylene.
  For excited states,
  ppTDA@RS-DFA provides accurate excitation energies and largely eliminates the DFA dependence.
  ppTDA@RS-DFA thus provides an efficient multireference approach to systems with static correlation.
\end{abstract}
% Opening: DFT and its limitations
Density functional theory (DFT)\cite{hohenbergInhomogeneousElectronGas1964,kohnSelfConsistentEquationsIncluding1965,parrDensityFunctionalTheoryAtoms1989}
has become the most widely used tool in the electronic structure theory.
The success of Kohn-Sham DFT (KS-DFT) can be attributed to the accurate
and efficient description of complicated many-body effects by density
functional approximations (DFAs), including local density approximations
(LDAs)\cite{barthLocalExchangecorrelationPotential1972,voskoAccurateSpindependentElectron1980}
, generalized gradient approximations (GGAs)\cite{beckeDensityfunctionalExchangeenergyApproximation1988,leeDevelopmentColleSalvettiCorrelationenergy1988,perdewAccurateSimpleAnalytic1992,perdewGeneralizedGradientApproximation1996}
and hybrid functionals\cite{beckeDensityfunctionalExchangeenergyApproximation1988,beckeNewMixingHartree1993}.
In past decades, KS-DFT has been successfully implemented in quantum
chemistry packages to predict properties of molecules and solids.
However, there are still many challenges in KS-DFT. For example, KS-DFT
cannot properly describe systems with a strong static correlation,
such as molecules with partially broken bonds\cite{zhangMulticonfigurationDensityCoherenceFunctional2021,limanniMulticonfigurationPairDensityFunctional2014,renRenormalizedSecondorderPerturbation2013}.
These strong correlated systems with degenerate or nearly degenerate
states are multiconfigurational and thus is challenging for the single-determinant-based
KS-DFT. It was shown that the errors of KS-DFT for predicting dissociation
energies of multiconfigurational systems are from fractional spin
errors in the functional approximation\cite{cohenInsightsCurrentLimitations2008,cohenFractionalSpinsStatic2008}.
To properly describe the bond breaking problem, multiconfiguration
self-consistent field (MCSCF) methods like the complete active space
self-consistent field (CASSCF) method\cite{roosCompleteActiveSpace1980,roosMulticonfigurationalMCSCF1983}
and other multireference methods in the wave function theory (WFT)
are commonly used. Extending the traditional Kohn-Sham DFT, multiconfigurational
methods have also been developed for describing multireference systems,
including multiconfigurational pair-density functional theory (MC-PDFT)\cite{gagliardiMulticonfigurationPairDensityFunctional2017,ghoshMulticonfigurationPairDensityFunctional2015,hoyerMulticonfigurationPairDensityFunctional2016}
and the multistate density functional theory (MSDFT)\cite{grofeSpinMultipletComponentsEnergy2017,renMultistateDensityFunctional2016,gaoKohnShamApproximation2016}.
Developments based on KS-DFT have also been made in past decades.
The fractional-spin localized orbital scaling correction (FSLOSC)\cite{suDescribingStrongCorrelation2018} method corrects fractional spin errors and fractional charge errors within the KS-DFT framework.
FSLOSC restores the flat-plane
behavior of electronic energy at fractional charges and fractional
spins and properly describes the dissociation of ionic species,
single bonds, and multiple bonds while maintaining the proper symmetry
in spin densities.
The spin-restricted ensemble-referenced Kohn-Sham (REKS) method\cite{filatovSpinrestrictedEnsemblereferencedKohn2015}
based on the rigorous ensemble representation of the energy and the density
provides accurate description for strongly correlated systems.
The commonly used time-dependent density functional theory\cite{ullrichTimeDependentDensityFunctionalTheory2011,rungeDensityFunctionalTheoryTimeDependent1984,casidaTimeDependentDensityFunctional1995} (TDDFT) is known
to provide poor accuracy for charge transfer and Rydberg excitations\cite{casidaProgressTimeDependentDensityFunctional2012,peachTripletInstabilityTDDFT2013}.
Recently developments including orbital optimized (OO) DFT\cite{haitOrbitalOptimizedDensity2021}
and mixed-reference spin-flip (MRSF)-TDDFT\cite{horbatenkoMixedReferenceSpinFlipTimeDependent2021}
are shown to provides considerable improvements over KS-DFT and TDDFT. \\

% review of multireference DFT
Recently a self-consistent multireference approach has been developed in DFT based
on linear response approaches \cite{chenMultireferenceDensityFunctional2017}.
In this multireference DFT approach,
the physical system is described by a generalized auxiliary system.
The total energy of the physical system is divided into two parts:
the energy of the auxiliary system
determined by the chosen DFA and the excitation energy from a linear
response theory. The excitation energy from the linear response theory
is used to provide multiconfigurational description for the physical
system in a natural way\cite{chenMultireferenceDensityFunctional2017}.
In the original work of the multireference DFT approach, two linear
response theories are used to provide the multiconfigurational description.
In the first approach,
the generalized auxiliary system is chosen to be the ($N-2$)-electron
system described with a DFA and the two-electron addition excitation
energy is calculated by particle-particle random phase approximation
(ppRPA) \cite{vanaggelenExchangecorrelationEnergyPairing2013,vanaggelenExchangecorrelationEnergyPairing2014,yangBenchmarkTestsSpin2013}
. The method is denoted here as ppRPA@DFA. This ppRPA@DFA approach
can also be viewed as a Fock-Space embedding approach that seamlessly
combines the many-body description of the two-electron subsystem with
the DFT description of the remaining ($N-2$) electrons\cite{zhangAccurateEfficientCalculation2016}.
The ppRPA method has also been explored recently by other research
groups\cite{bannwarthHoleHoleTamm2020,tahirComparingParticleparticleParticlehole2019}.
In the second approach,
the generalized auxiliary system is chosen to be a high-spin state.
Then low-spin states are obtained by using spin-flip time-dependent
DFT (SF-TDDFT)\cite{shaoSpinFlipApproach2003,wangTimedependentDensityFunctional2004,bernardGeneralFormulationSpinflip2012,liTheoreticalNumericalAssessments2012}.
This approach is denoted here as SF@DFA. In both cases, the total
energy in the multireference DFT approach is an implicit functional
of the density of the physical system. The electron density of the
physical systems is given by the functional derivative of the total
energy with respect to the external field\cite{chenMultireferenceDensityFunctional2017}.
It is clearly not equal to the density of the auxiliary reference
system, in contrast to the traditional KS theory\cite{jinIntroductoryLectureWhen2020}.
The self-consistent field (SCF) solution, or the minimization of the
energy functional was obtained by the generalized optimized effective
potential (GOEP) method\cite{chenMultireferenceDensityFunctional2017,jinGeneralizedOptimizedEffective2017}.
It has been shown that the self-consistency is important for ppRPA@DFA
and SF@DFA to predict more accurate dissociation energies than the
post-SCF approaches: post-SCF-ppRPA@DFA and post-SCF-SF@FDA\cite{chenMultireferenceDensityFunctional2017}.
Here post-SCF means the total energy in the multireference DFT approach is
evaluated without the self-consistency.
\\

% Challenge in multireference DFT
The multireference DFT approach was shown to provide good descriptions
for both ground and excited states\cite{chenMultireferenceDensityFunctional2017}.
The ppRPA@HF approach properly describes bond breaking of small molecules
and the the double bond rotation of the $\text{C}_{2}\text{H}_{4}$
molecule. The SF@HF provides accurate excitation energies of atomic
and molecular systems. However, there are two problems in the original multireference
DFT approach. First, as demonstrated in this work, the multireference
DFT approach with commonly used DFAs has large errors for predicting
energies of both ground states and excited states. The accuracy of
the multireference DFT approach has an undesired dependence on the
choice of the DFA. Second, the optimization of the GOEP method for
multireference DFT with traditional DFAs can produce an unphysical
density of the generalized auxiliary system, which has a negative
HOMO-LUMO gap, or the switching of the HOMO and LUMO. This unphysical
density leads to a failed optimization. Thus, only HF and a designed
DFA with a high percentage of exact exchange were used in the original
work\cite{chenMultireferenceDensityFunctional2017}. The orbital optimization
(OO)\cite{head-gordonOptimizationWaveFunction1988,peveratiOrbitalOptimizedDoublehybrid2013,yaffeOrbitalOptimizationElectronic1976}
can also be developed for the energy optimization in multireference
DFT, as shown in the Section.S1 of the Supporting Information. Although
OO has a better convergence rate than GOEP, it is shown to be equivalent
to the optimization of an orbital functional with respect to the GOEP
Hamiltonian\cite{jinGeneralizedOptimizedEffective2017} and is thus
expected to have the similar problem as with the GOEP method. Not
using the optimal DFA can lead to large error in energies of the generalized
auxiliary systems, for example, for the (N-2)-electron system in the
case of the ppRPA@DFA approach. Ideally, one would like to use the
optimal DFA to describe the (N-2)-electron reference system, because
its energy contributes to the major part of the total energy for the
N-electron system.\\

% Resolution 1: RS
In the present work, to overcome the first challenge, we applied renormalized
singles (RS) in the multireference DFT. RS was developed to provide
a good starting point in the $GW$ calculations\cite{jinRenormalizedSinglesGreen2019},
which is the RS Green's function. The motivation of the RS Green's
function is to use the form of the Hartree-Fock\cite{szaboModernQuantumChemistry2012,slaterNoteHartreeMethod1930}
(HF) self-energy to include all singles contributions. Because of
the Brillouin theorem\cite{szaboModernQuantumChemistry2012}, the
HF self-energy captures all the contributions of the single excitations
completely. However, the HF Green's function itself is not a good
starting point for $GW$ calculations\cite{jinRenormalizedSinglesGreen2019}.
Therefore, the HF Hamiltonian is constructed with KS orbitals and
diagonalized separately in the occupied and virtual subspaces of the
DFA used. This renormalization procedure absorbs all singles contributions
into the self-energy to eliminate the dependence on the choice of
the DFA. The RS Green's function constructed with RS eigenvalues is
shown to predict accurate quasiparticle energies for valence states
in the $GW$ method as well as valence and core states in the T-matrix
methods\cite{jinRenormalizedSinglesGreen2019,liRenormalizedSinglesGreen2021}.
The RS Green's function shares a similar thinking as the renormalized
single-excitation (rSE) correction in the random phase approximation
(RPA) calculations for the correlation energy\cite{renRandomPhaseApproximationElectron2011,renRenormalizedSecondorderPerturbation2013}.
The accuracy of RPA with rSE calculations for predicting binding energies
of rare-gas dimers is significantly improved\cite{renRandomPhaseApproximationElectron2011,renRenormalizedSecondorderPerturbation2013}.
In our work, RS changes the linear response calculation of the multireference
DFT approach for ground and excited states, and the renormalization
process does not modify the one-electron density matrix of the ($N-2$)-electron
auxiliary system and consequently its energy from a given DFA. The
renormalization process that includes contributions of all one-body
perturbations from the auxiliary system is expected to bring a better
description for the linear response property and the resulting multireference
description for the static correlation, as RS does for improving significantly
the $GW$ method and the T-matrix method\cite{jinRenormalizedSinglesGreen2019,liRenormalizedSinglesGreen2021}.
Based on the ($N-2$)-electron system as the auxiliary system,
the two-electron addition energy is obtained from particle-particle
Tamm-Dancoff approximation\cite{yangExcitationEnergiesParticleparticle2014,yangDoubleRydbergCharge2013}
(ppTDA) in the present work, to simplify the calculations without
lost of accuracy\cite{yangExcitationEnergiesParticleparticle2014}.
The RS Hamiltonian is used in the working equation of ppTDA. We denote
this approach as ppTDA@RS-DFA. The corresponding post-SCF approach
that calculates the total energy without the self-consistency is denoted
as post-SCF-ppTDA@RS-DFA. \\

% Resolution 2: OEP
To address the second issue on the failure of the GOEP optimization
for multireference DFT when applied with commonly used DFAs, we introduced
the optimized effective potential (OEP) method of Yang and Wu\cite{yangDirectMethodOptimized2002}
for the total energy optimization in the multireference DFT approach,
in contrast to the GOEP method used in the original work\cite{chenMultireferenceDensityFunctional2017}.
The Yang and Wu OEP method is an accurate and efficient method for
the direct energy optimization for functionals depending on one-particle
density matrix. A solution to the inverse problem of calculating a
set of orbitals and the density matrix from a given electron density
has also been developed similarly\cite{wuDirectOptimizationMethod2003}.
In the OEP method, the local potential is constructed from the sum
of a fixed one-electron potential and a linear combination of potential
basis functions. This turns the difficult-to-solve integral equation
in OEP to a direct optimization with respect to the coefficients of
the potential basis functions, with the correct asymptotic conditions
of the potential enforced by the long-range fixed one-electron potential.
The OEP method has been implemented for calculating total energies,
potentials and geometries for a broad range of systems\cite{yangDirectMethodOptimized2002,wuDirectOptimizationMethod2003,wuAnalyticEnergyGradients2005}.
In this paper, we explored the optimization of multireference DFT
with the OEP method. The gradient of the total energy with respect
to the coefficients of the linear expansion in multireference DFT
consists of two parts. The gradient of the ($N-2$)-electron auxiliary
system with respect to the potential basis function only involves
the ($N-2$)-electron Fock matrix. And the gradient of the ppTDA excitation
energies is formulated with eigenvectors of ppTDA and the exchange-correlation
(XC) kernel of the chosen DFA, which was derived in previous works\cite{chenMultireferenceDensityFunctional2017,jinGeneralizedOptimizedEffective2017}.
We demonstrate that the optimization of the multireference DFT approach
with conventional DFAs in the OEP method can produce physical potentials
for most systems. In contrast, the GOEP method fails to converge for
commonly used DFAs\cite{chenMultireferenceDensityFunctional2017}.
Our approach thus opens up broad possibilities for the application
of the multireference DFT approach. \\

% Actions
In the multireference DFT approach, the total energy of the $n-$th
eigenstate of the physical system $E_{v}[\rho]$ is expressed as\cite{chenMultireferenceDensityFunctional2017}
\begin{equation}
E_{v,n}[\rho]=E_{v}^{\text{ref}}[\rho_{s}^{\text{ref}}]+\Delta E_{n}[\rho_{s}^{\text{ref}}]\text{,}\label{eq:multiref_energy}
\end{equation}
where $\rho$ is the density matrix of the physical system, $\rho_{s}^{\text{ref}}$
is the density matrix of the noninteracting auxiliary reference system,
$E_{v}^{\text{ref}}[\rho_{s}^{\text{ref}}]$ is the energy of the
auxiliary system determined by the chosen DFA, $\Delta E_{n}[\rho_{s}^{\text{ref}}]$
is the excitation energy from a linear response theory that restores
the symmetry of the physical system, and $n$ is the energy level.
Eq.\ref{eq:multiref_energy} indicates that the energy of excited
states can also be obtained by selecting the target excitation energy
from the linear response theory. The noninteracting auxiliary reference
system can be an ($N-2$)-electron system in conjunction with a ppRPA
linear response for excitation or an $N$-electron high spin state
system in conjunction with a SF-TDDFT calculation for excitation\cite{chenMultireferenceDensityFunctional2017,jinGeneralizedOptimizedEffective2017}.
The density of the physical system $\rho$, not equal to $\rho_{s}^{\text{ref}}$,
is defined by the linear response of the total energy.\\

In present work, we chose the ($N-2$)-electron system as the auxiliary
system and the excitation energies are obtained by ppTDA for Eq.\ref{eq:multiref_energy}.
The working equation of ppTDA is obtained by keeping only the two-electron
addition part in the ppRPA matrix\cite{yangDoubleRydbergCharge2013,yangExcitationEnergiesParticleparticle2014,zhangAnalyticGradientsGeometry2014}
\begin{equation}
AX=\omega X\text{,}\label{eq:pptda}
\end{equation}
where
\begin{align}
A_{ab,cd} & =\delta_{ac}F_{bd}+\delta_{bd}F_{ac}+\langle ab||cd\rangle\text{,}\label{eq:a_matrix}
\end{align}
$F$ is the one-electron KS or generalized KS Hamiltonian of the DFA
used, $X$ is the two-electron addition
eigenvector, and $\omega$ is the two-electron addition energy. In
above equations, the antisymmetrized two-electron integral $\langle pq||rs\rangle$
is defined as
\begin{equation}
 \langle pq||rs\rangle=\langle pq|rs\rangle-\langle qp|rs\rangle\\
 =\int dx_{1}dx_{2}\frac{\phi_{p}^{*}(x_{1})\phi_{q}^{*}(x_{2})(1-\hat{P}_{12})\phi_{r}(x_{1})\phi_{s}(x_{2})}{|\mathbf{r_{1}}-\mathbf{r_{2}}|}
\end{equation}
where $x$ stands for both spatial and spin coordinates\cite{martinInteractingElectrons2016}.
We use $i$, $j$ for occupied orbitals, $a$, $b$ for virtual orbitals
and $p$, $q$ for general orbitals. Because we are only interested
in the ground state and low-lying excited states, Eq.\ref{eq:pptda}
can be efficiently solved by the Davidson algorithm\cite{davidsonIterativeCalculationFew1975,stratmannEfficientImplementationTimedependent1998}
as implemented in Ref.\citenum{yangExcitationEnergiesParticleparticle2014}.\\

Although Eq.\ref{eq:multiref_energy} provides a rigorous description
for ground states and excited states, its accuracy strongly depends
on the choice of the DFA. To reduce the dependence and improve the
accuracy, we now introduce the RS method. The RS Hamiltonian is defined
as\cite{jinRenormalizedSinglesGreen2019}
\begin{equation}
H^{\text{RS}}=PH^{\text{HF}}[\rho_{s}^{\text{ref}}]P+QH^{\text{HF}}[\rho_{s}^{\text{ref}}]Q\text{,}\label{eq:rs_fock}
\end{equation}
where $P=\sum_{i}^{occ}|\phi_{i}\rangle\langle\phi_{i}|$ is the projection
into the occupied orbital space and $Q=I-P$ is the projection into
the virtual orbital space defined by the DFA used, $H^{\text{HF}}[\rho_{s}^{\text{ref}}]$
means the HF Hamiltonian is constructed with the KS density matrix.
The RS Hamiltonian captures all contributions from single excitations
of the ($N-2$)-electron system. The RS Hamiltonian is inserted in
Eq.\ref{eq:pptda} as the generalized KS Hamiltonian of RS-DFA
to get the excitation energy, which is expected
to provide a better description for states with a multireference nature
and with the static correlation. Because the working equation of ppTDA
in Eq.\ref{eq:pptda} only has the two-electron addition part, only
the virtual-virtual block of the RS Hamiltonian is needed.\\

To achieve a SCF solution, we optimized the total energy in Eq.\ref{eq:multiref_energy}
with the OEP method. In the OEP method, the potential is constructed
as following\cite{yangDirectMethodOptimized2002}
\begin{equation}
v_{s}(\mathbf{r})=v_{\text{{ext}}}(\mathbf{r})+v_{0}(\mathbf{r})+\sum_{t}b_{t}g_{t}(\mathbf{r})\text{,}\label{eq:oep}
\end{equation}
where $v_{\text{{ext}}}(r)$ is the external potential due to the
nuclei, $v_{0}(r)$ is a fixed potential and $\{b_{t}\}$ are the
coefficients for the potential basis sets $\{g_{t}\}$. The Fermi-Amaldi
potential is chosen as the fixed potential $v_{0}(r)$\cite{zhaoElectronDensitiesKohnSham1994},
which is defined as
\begin{equation}
v_{0}(\mathbf{r})=\frac{N_{0}-1}{N_{0}}\int\frac{\rho_{0}(\mathbf{r})}{|\mathbf{r}-\mathbf{r'}|}d\mathbf{r'}\text{,}\label{eq:fa_potential}
\end{equation}
where $N_{0}$ is the number of electrons corresponding to a fixed
density $\rho_{0}$. The optimization with respect to the coefficients
$\{b_{t}\}$ can be easily performed with gradient-based approaches.
In principle, because of the flexibility from the last term in Eq.\ref{eq:oep},
the fixed density $\rho_{0}$ in Eq.\ref{eq:fa_potential} can be
arbitrary, as long as the potential basis set $\{g_{t}\}$ is sufficient.
The fixed potential $v_{0}(r)$ using the density from the SCF calculation
of the ($N-2$)-electron system or the $N$-electron system give similar
results as shown in the Section.S3 of the Supporting Information.
In practice, we found that the density from the SCF calculation of
the ($N-2$)-electron system is a better choice to produce physical
potentials and give better convergence behaviors.
We thus use it in Eq.\ref{eq:fa_potential} throughout this work.\\

Following Ref.\citenum{yangDirectMethodOptimized2002}, the gradient
of the total energy in multireference DFT with respect to the coefficients
$\{b_{t}\}$ consists of two parts: the auxiliary system energy and
the excitation energy
\begin{equation}
\frac{\partial E_{v,n}}{\partial b_{t}}=\sum_{i,a}\bigg(\langle\frac{\delta E_{v}^{\text{ref}}}{\delta\phi_{i}}|\phi_{a}\rangle+\langle\frac{\delta\Delta E_{n}}{\delta\phi_{i}}|\phi_{a}\rangle\bigg)\frac{\langle\phi_{a}|g_{t}|\phi_{i}\rangle}{\epsilon_{i}-\epsilon_{a}}+\text{c.c.}\label{eq:gradient}
\end{equation}
where $\epsilon_{i}$ and $\epsilon_{a}$ are the occupied and the
virtual orbital energy. The gradient of the ($N-2$)-electron auxiliary
system energy only involves the matrix elements of the(generalized)
Kohn-Sham Hamiltonian $H^{\text{ref}}$\cite{chenMultireferenceDensityFunctional2017}
\begin{equation}
\frac{\partial E_{v,n}}{\partial b_{t}}=-\sum_{i,a}H^{\text{ref}}_{ia}\frac{\langle\phi_{a}|g_{t}|\phi_{i}\rangle}{\epsilon_{i}-\epsilon_{a}}\text{.}\label{eq:gradient_dfa}
\end{equation}
Second, the gradient of the excitation energy from ppTDA is formulated
as\cite{chenMultireferenceDensityFunctional2017}
\begin{equation}
\frac{\partial\Delta E_{v,n}}{\partial b_{t}}=\sum_{i,a}\frac{\langle\phi_{a}|g_{t}|\phi_{i}\rangle}{\epsilon_{i}-\epsilon_{a}}\bigg\{2\sum_{a'b'}X_{aa'}^{n}H_{ib'}X_{b'a'}^{n}+\sum_{a'b'c'}X_{aa'}^{n}\langle ia'||b'c'\rangle X_{b'c'}^{n}-2\sum_{a'b'c'}X_{a'b'}^{n}K_{a'c',ai}^{\text{Hxc}}X_{c'b'}^{n}\bigg\}\text{.}\label{eq:gradient_pp}
\end{equation}
where $X_{ab}^{n}$ is the eigenvector of the ppTDA equation defined
in Eq.\ref{eq:pptda}, $K^{\text{Hxc}}$ is the Hartree-exchange-correlation
(Hxc) kernel that is the second derivative of the Hxc energy with
respect to the electron density. In our ppTDA@RS-DFA approach, because
the RS Hamiltonian defined in Eq.\ref{eq:rs_fock} has the same form
as the HF matrix, the evaluation of the kernel $K^{\text{Hxc}}$ only
involves two-electron integrals, which is
\begin{equation}
K_{pq,rs}^{\text{RS,Hxc}}=\langle qp||rs\rangle\label{eq:rs_kernel}
\end{equation}
This leads to better efficiency in multireference DFT calculations for molecules.

In practical calculations, the frontier orbitals of the ($N-2$)-electron
auxiliary system for a given trial potential $v_{s}(\mathbf{r}),$
can be degenerate or nearly degenerate, which gives incorrect gradient
in Eq.\ref{eq:gradient}.
Numerical techniques in the many-body perturbation theory such as using an
energy level shift in the denominator in Eq.\ref{eq:gradient} lead to the
wrong gradient and failed optimization.
To address the degeneracy in the optimization,
we developed a hybrid algorithm. When the frontier orbitals are not
nearly degenerate, the Broyden-Fletcher-Goldfarb-Shanno\cite{pressNumericalRecipes2nd1992}
(BFGS) method is used. When the HOMO and LUMO orbitals, or more of
the frontier orbitals of the ($N-2$)-electron auxiliary system are
degenerate, its single determinant wavefunction is not uniquely defined.
In addition, the gradient cannot be calculated from Eq.\ref{eq:gradient}
because of the singularity arising from the small HOMO-LUMO gap in
the denominators. In case of degenerate frontier orbitals, we first
selected the unique determinant wavefunction, $\Psi_{v}$, or the
unique set of $(N-2)$ orbitals by the minimization of the total energy
functional with respect to the selected frontier orbitals defined
as unitary rotations of all frontier orbitals in the degenerate subspace.
In the simple case of only degenerate HOMO and LUMO,
the unitary rotation is only performed between two frontier orbitals,
this being the most frequent cases when our calculations
encountered degeneracy. Thus the unique determinant wavefunction $\Psi_{v}$
is given as
\begin{equation}
\Psi_{v}=arg\min_{\tilde{\Psi}}E_{v,n}[\tilde{\Psi}]\text{.}\label{eq:optimal_det}
\end{equation}
where $\tilde{\Psi}$ is the trial determinant for the ($N-2$)-electron
auxiliary system with its frontier orbitals defined by the unitary
rotation in the degenerate subspace and $E_{v,n}[\tilde{\Psi}]$ is
the total energy functional of Eq. \ref{eq:multiref_energy} but with
the change of the variable from the density matrix to the determinant
wavefunction to emphasize the determinant nature of the solution needed
for the ppRPA or ppTDA linear response calculations.
Thus only orbitals in the degenerate/nearly degenerate subspace are mixed.
A single determinant $\Psi_{v}$ is thus uniquely defined for a trial
potential $v_{s}(\mathbf{r})$ with degenerate frontier orbitals.
This is a new development and a departure from the original formulation
of potential functional theory (PFT)\cite{yangPotentialFunctionalsDual2004},
the theoretical foundation for the OEP approaches, where ensemble
solutions have been used. \\

The optimization of energy with respect to the unitary transformation
of degenerate or nearly degenerate frontier orbitals in the determinant
wavefunction in Eq. \ref{eq:optimal_det}
\begin{equation}
    U=\prod_{p\in \text{deg}}\bigg[\prod_{q\in\text{deg},q<p}G_{pq}(\theta_{pq})\bigg]
\end{equation}
is carried out with the sequential rotations. Here deg means degenerate
or nearly degenerate subspace, $G_{pq}$ is the $2\times2$ rotation
matrix between orbital $p$ and $q$, $\theta_{pq}$ is the rotation
angle. In the optimization of each orbital pair, the rotation angle
is determined by a three-point parabolic interpolation. The sequential
rotations are performed in a consistent ordering until the energy
minimum is reached.\\

With the single determinant wavefunction uniquely defined for the
trial potential with degenerate frontier orbitals, we next consider
the optimization step for the trial potential. In such a degenerate
case, the functional derivative of the total energy with respect to
the potential is not defined; only the energy variation for a given
change of the potential is defined. Thus the energy gradient of the
optimization defined in Eq.\ref{eq:optimal_det} is not defined as
shown in the Section.S4 of Supporting Information. One way forward
is to carry out optimization of potential without gradients.
Instead, we use an approximate gradient with respect to the potential.
The approximate gradient has a similar form as Eq.\ref{eq:gradient}
without contributions from degenerate or nearly degenerate orbitals
\begin{equation}
\frac{\partial E_{v,n}}{\partial b_{t}}=\sum_{i,a\notin\text{deg}}\bigg(\langle\frac{\delta E_{v,n}^{\text{ref}}}{\delta\phi_{i}}|\phi_{a}\rangle+\langle\frac{\delta\Delta E_{v,n}}{\delta\phi_{i}}|\phi_{a}\rangle\bigg)\frac{\langle\phi_{a}|g_{t}|\phi_{i}\rangle}{\epsilon_{i}-\epsilon_{a}}+\text{c.c.}\label{eq:approx_gradient}
\end{equation}
A line search along the approximate gradient defined in Eq.\ref{eq:approx_gradient}
is then performed to obtain a new total energy and a new trial potential.
This hybrid algorithm appropriately deals with the accidental degenerate
cases in the optimization of the OEP method.
In this work,
we show that the ppTDA@RS-DFA approach provides the converged results in
most ($N-2$)-electron systems,
which are difficult to obtain in the conventional SCF calculation\cite{yangExcitationEnergiesParticleparticle2014}.
For degeneracy originated from the symmetry of systems,
the degeneracy persists during the optimization of the potential and
the gradient can not be well approximated by Eq.\ref{eq:approx_gradient}.
Further study, such as optimization without gradient, is needed for such cases.\\

% Computational details
We implemented the ppTDA@DFA approach and the ppTDA@RS-DFA approach
in the QM4D quantum chemistry package\cite{qm4d} and applied the method
to calculate single bond dissociations, double bond rotations and
excitation energies of molecular systems. For the single bond dissociation,
the cc-pVTZ basis set\cite{dunningGaussianBasisSets1989} was used
for LiH, BH and $\text{CH}^{+}$, while the cc-pVDZ basis set\cite{dunningGaussianBasisSets1989}
was used for $\text{C}_{2}\text{H}_{6}$ considering the computational
cost. In the dissociation of $\text{C}_{2}\text{H}_{6}$, the C-C
bond was stretched. The equilibrium geometry of $\text{C}_{2}\text{H}_{6}$
was taken from Ref.\citenum{liDrivenSimilarityRenormalization2017}.
In ppTDA@DFA, ppTDA@RS-DFA, CCSD calculations, the total energies at
the bond length 5.0Å, 6.0Å, 4.0Å, 5.0Å were set as zero for LiH, BH,
$\text{CH}^{+}$, $\text{C}_{2}\text{H}_{4}$ respectively. In KS-DFT
calculations, the total energies of the doublet Li atom and the doublet
H atom, the doublet B atom and the doublet H atom, the doublet $\text{C}^{+}$
cation and the doublet H atom, two double C$\text{H}_{3}$ doublets
were set as zero for LiH, BH, $\text{CH}^{+}$, and $\text{C}_{2}\text{H}_{6}$,
respectively.
The reference for LiH was taken from Ref.\citenum{liAccurateDeterminationRovibrational2003},
for BH was taken from Ref.\citenum{abramsAssessmentAccuracyMultireference2003},
for CH$^{+}$ was taken from Ref.\citenum{biglariInitioPotentialEnergy2014},
for $\text{C}_{2}\text{H}_{6}$ was taken from Ref.\citenum{liDrivenSimilarityRenormalization2017}.
For the double bond rotation of $\text{C}_{2}\text{H}_{4}$,
the cc-pVTZ basis set was used. Geometries of $\text{C}_{2}\text{H}_{4}$
were taken from Ref.\citenum{chenMultireferenceDensityFunctional2017}.
The reference MR-ccCA values were taken from Ref.\citenum{jiangEmpiricalCorrectionNondynamical2012}.
For excitation energies, the cc-pVTZ basis set was used for LiH and
Ca, the aug-cc-pVDZ basis set was used for $\text{C}_{2}\text{H}_{4}$,
and the aug-cc-pTVZ basis set was used for remaining systems. Geometries
of BH, $\text{CH}^{+}$, CO and $\text{C}_{2}\text{H}_{4}$ were taken
from Ref.\citenum{chenMultireferenceDensityFunctional2017}. The
geometry of $\text{H}_{2}$O was taken from CORE65 set\cite{golzeAccurateAbsoluteRelative2020}.
The geometry of LiH was optimized at the CCSD level with cc-pVTZ
basis set. The reference value of LiH was obtained from EOM-CCSD with
the cc-pVTZ basis set. The reference value of $\text{H}_{2}$O was
taken from Ref.\citenum{hoyerMulticonfigurationPairDensityFunctional2016}.
Reference values of remaining systems were taken from Ref.\citenum{chenMultireferenceDensityFunctional2017}.
ppTDA@DFA, ppTDA@RS-DFA, post-SCF-ppTDA@DFA and post-SCF-ppTDA@RS-DFA calculations
were performed with QM4D, KS-DFT, CCSD, TDDFT and EOM-CCSD calculations were
performed with GAUSSIAN16 A.03 software\cite{g16}. QM4D uses Cartesian
basis sets and the resolution of identity (RI) technique\cite{weigendAccurateCoulombfittingBasis2006,renResolutionofidentityApproachHartree2012,eichkornAuxiliaryBasisSets1995}
to compute two-electron integrals in multireference DFT calculations.
All basis sets were taken from the Basis Set Exchange\cite{fellerRoleDatabasesSupport1996,pressNumericalRecipes2nd1992,schuchardtBasisSetExchange2007}.

\begin{figure*}
\centering \begin{subfigure}[b]{0.475\textwidth} \centering
\includegraphics[width=1\textwidth]{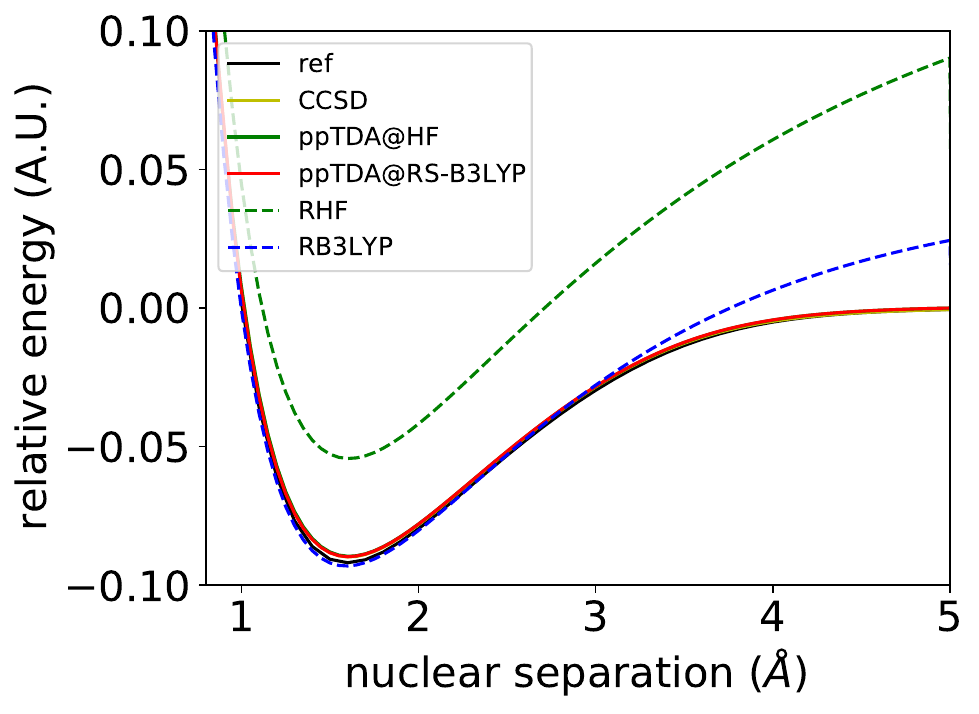} \caption{LiH}
\label{fig:curve_lih} \end{subfigure} \hfill{}\begin{subfigure}[b]{0.475\textwidth}
\centering \includegraphics[width=1\textwidth]{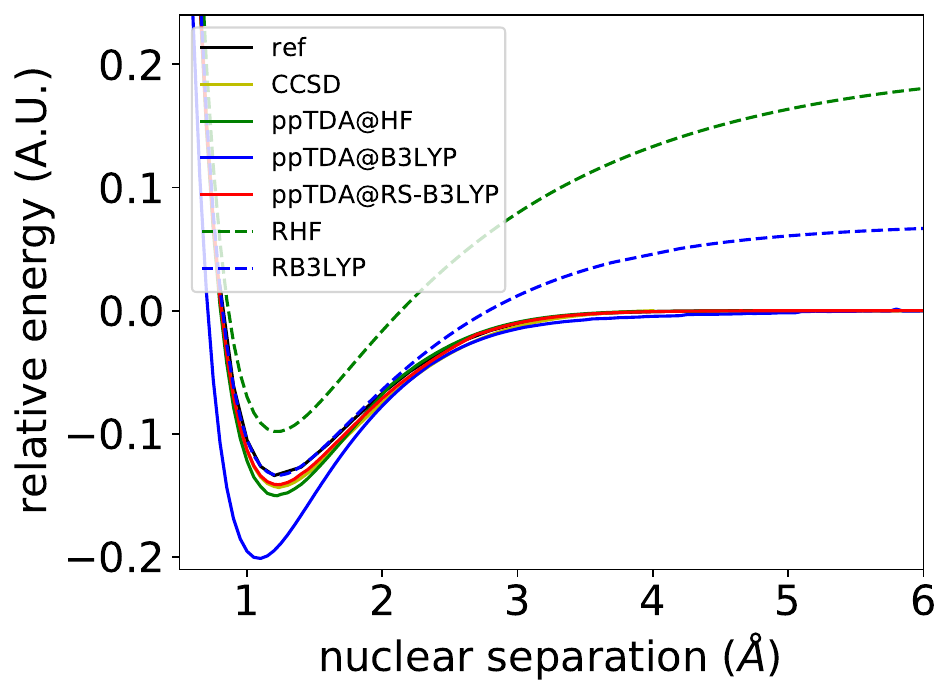}
\caption{BH}
\label{fig:curve_bh} \end{subfigure} \vskip\baselineskip \begin{subfigure}[b]{0.475\textwidth}
\centering \includegraphics[width=1\textwidth]{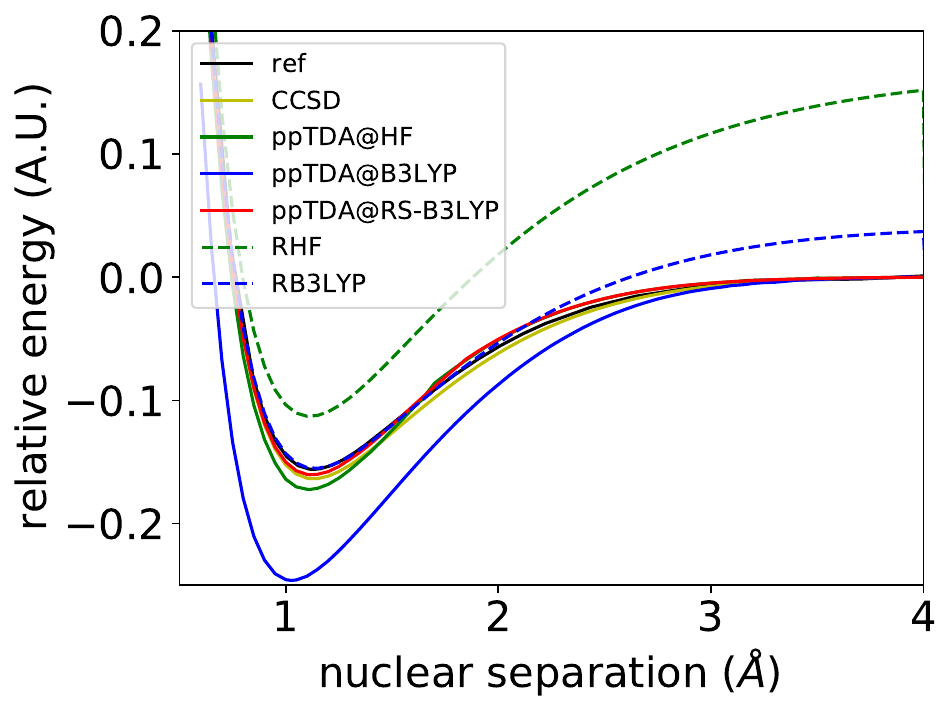}
\caption{$\text{CH}^{+}$}
\label{fig:curve_chp} \end{subfigure} \hfill{}\begin{subfigure}[b]{0.475\textwidth}
\centering \includegraphics[width=1\textwidth]{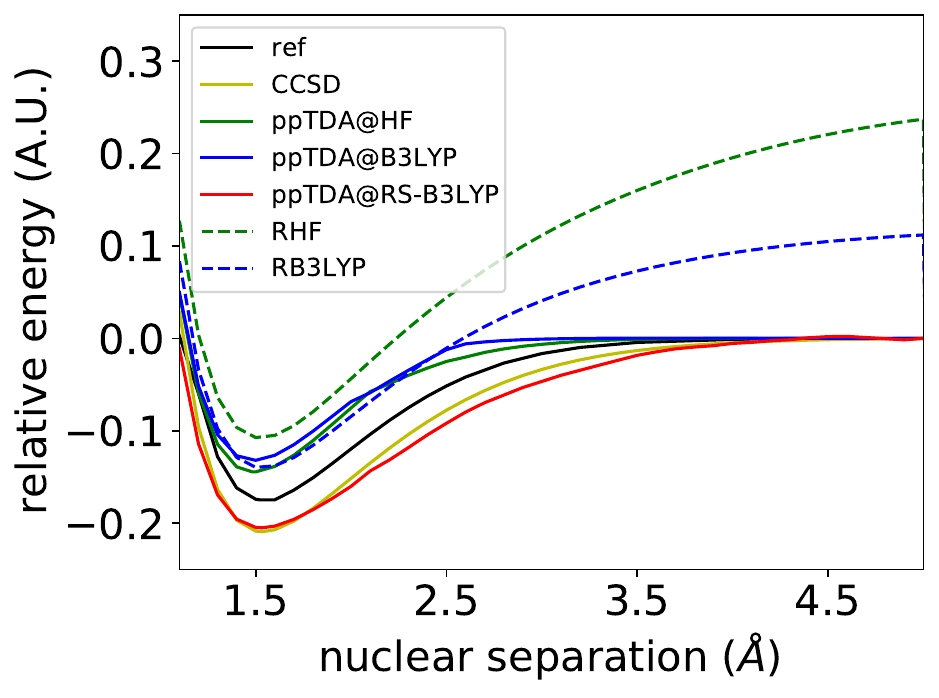}
\caption{$\text{C}_{2}\text{H}_{6}$}
\label{fig:curve_c2h6} \end{subfigure} \caption{Dissociation curves of LiH, BH, $\text{CH}^{+}$ and $\text{C}_{2}\text{H}_{6}$
obtained from ppTDA@HF, ppTDA@B3LYP, ppTDA@RS-B3LYP, restricted HF
(RHF) and restricted B3LYP (RB3LYP) calculations. The cc-pVTZ basis set was used for LiH, BH and
$\text{CH}^{+}$. The cc-pVDZ basis set was used for $\text{C}_{2}\text{H}_{6}$.
Reference for LiH was taken from Ref.\citenum{liAccurateDeterminationRovibrational2003},
for BH was taken from Ref.\citenum{abramsAssessmentAccuracyMultireference2003},
for CH$^{+}$ was taken from Ref.\citenum{biglariInitioPotentialEnergy2014},
for $\text{C}_{2}\text{H}_{6}$ was taken from Ref.\citenum{liDrivenSimilarityRenormalization2017}.}
\label{fig:dissociation_curves}
\end{figure*}

% Results: dissociation energy
We first examined the performance of the ppTDA@RS-DFA approach for
describing single bond breakings. The bond dissociation curves of
LiH, BH, $\text{CH}^{+}$ and $\text{C}_{2}\text{H}_{6}$ molecules
are shown in Fig.\ref{fig:dissociation_curves} with tabulated results
of dissociation energies and equilibrium bond lengths shown in Table.\ref{tab:dissociation_energy}
and Table.\ref{tab:bond_length}. The results show that the ppTDA@RS-DFA
approach describes well the single bond dissociation process while
preserving the proper spin density symmetry of the solution and
significantly outperforms ppTDA@DFA and restricted KS-DFT
for predicting both dissociation energies and equilibrium bond lengths.
It is well known that the single bond dissociation cannot be properly
described by single-determinant KS-DFT calculations with commonly
used DFAs with proper spin density symmetry (in restricted calculations).
Although restricted B3LYP and restricted HF give reasonable equilibrium bond lengths, they produce too high energies at dissociation limits,
leading to large errors for dissociation energies.
Compared with KS-DFT, the ppTDA@DFA approach
provides correct dissociation limits for all systems because of proper
multiconfigurational descriptions for broken bonds. However, the accuracy
for predicting dissociation energies is not satisfactory. ppTDA@HF
and ppTDA@B3LYP have a mean absolute error (MAE) for the dissociation
energy of $0.45$ \,{eV} and $1.99$ \,{eV}, respectively. By
including all one-body perturbations from the auxiliary system, our
ppTDA@RS-B3LYP approach predicts accurate dissociation energies and
equilibrium bond lengths for all systems.
The ppTDA@RS-DFA approach even provides more accurate dissociation energies
than the computationally expensive CCSD approach.
The MAE of ppTDA@RS-B3LYP for predicting
dissociation energies is only $0.30$ \,{eV} and for predicting
equilibrium bond lengths is only $0.01$ Å. \\

\begin{table}
\caption{Dissociation energies (eV) of LiH, BH, $\text{CH}^{+}$ and $\text{C}_{2}\text{H}_{6}$
obtained from restricted HF, restricted B3LYP, ppTDA@HF, ppTDA@B3LYP
and ppTDA@RS-B3LYP.}
\label{tab:dissociation_energy}\centering \begingroup \begin{threeparttable}
\global\long\def\arraystretch{1.5}%
% Default value: 1
\setlength{\tabcolsep}{15pt} %
\begin{tabular}{cccccc}
\hline
 & LiH\tnote{1}  & BH  & $\text{CH}^{+}$  & $\text{C}_{2}\text{H}_{6}$  & MAE\tnote{2}\tabularnewline
 ref              & 2.50  & 3.64  & 4.24  & 4.75  &      \tabularnewline
CCSD              & 2.44  & 3.91  & 4.45  & 5.69  & 0.37 \tabularnewline
\multicolumn{6}{c}{KS-DFT}\tabularnewline
restricted HF     & 3.93  & 7.47  & 7.20  & 9.39  & 1.25 \tabularnewline
restricted B3LYP  & 3.20  & 5.32  &       & 6.86  & 0.25 \tabularnewline
\multicolumn{6}{c}{Multireference DFT}\tabularnewline
ppTDA@HF          & 2.43  & 4.09  & 4.69  & 3.93  & 0.45 \tabularnewline
ppTDA@B3LYP       &       & 5.47  & 6.70  & 3.59  & 1.99 \tabularnewline
ppTDA@RS-B3LYP    & 2.44  & 3.84  & 4.36  & 5.58  & 0.30 \tabularnewline
\hline
\end{tabular}

\begin{tablenotes}

  \item{[}1{]} ppTDA@B3LYP for LiH result is missing because the failure
  of the optimization.

  \item{[}2{]} MAE stands for mean absolute error.

  \item{[}3{]} The reference for LiH was taken from Ref.\citenum{liAccurateDeterminationRovibrational2003},
  for BH was taken from Ref.\citenum{abramsAssessmentAccuracyMultireference2003},
  for CH$^{+}$ was taken from Ref.\citenum{biglariInitioPotentialEnergy2014},
  for $\text{C}_{2}\text{H}_{6}$ was taken from Ref.\citenum{liDrivenSimilarityRenormalization2017}.
\end{tablenotes}
\end{threeparttable} \endgroup
\end{table}

\begin{table}
\caption{Equilibrium bond lengths (Å) of LiH, BH, $\text{CH}^{+}$ and $\text{C}_{2}\text{H}_{6}$
obtained from restricted HF, restricted B3LYP, ppTDA@HF, ppTDA@B3LYP
and ppTDA@RS-B3LYP.}
\label{tab:bond_length}\centering \begingroup \begin{threeparttable}
\global\long\def\arraystretch{1.5}%
% Default value: 1
 \setlength{\tabcolsep}{15pt} %
\begin{tabular}{cccccc}
\hline
 & LiH\tnote{1}  & BH  & $\text{CH}^{+}$  & $\text{C}_{2}\text{H}_{6}$  & MAE\tnote{2} \tabularnewline
ref               & 1.60  & 1.20  & 1.12  & 1.52  &      \tabularnewline
CCSD              & 1.61  & 1.24  & 1.11  & 1.53  & 0.02 \tabularnewline
\multicolumn{5}{c}{KS-DFT} & \tabularnewline
restricted HF     & 1.61  & 1.22  & 1.11  & 1.53  & 0.01 \tabularnewline
restricted B3LYP  & 1.59  & 1.23  &       & 1.53  & 0.02 \tabularnewline
\multicolumn{5}{c}{Multireference DFT} & \tabularnewline
ppTDA@HF          & 1.61  & 1.22  & 1.11  & 1.48  & 0.02 \tabularnewline
ppTDA@B3LYP       &       & 1.09  & 1.03  & 1.50  & 0.46 \tabularnewline
ppTDA@RS-B3LYP    & 1.61  & 1.23  & 1.11  & 1.53  & 0.01 \tabularnewline
\hline
\end{tabular}
\begin{tablenotes}

    \item{[}1{]} ppTDA@B3LYP for LiH result is missing because the failure
    of the optimization.

    \item{[}2{]} MAE stands for mean absolute error.

    \item{[}3{]} The reference for LiH was taken from Ref.\citenum{liAccurateDeterminationRovibrational2003},
    for BH was taken from Ref.\citenum{abramsAssessmentAccuracyMultireference2003},
    for CH$^{+}$ was taken from Ref.\citenum{biglariInitioPotentialEnergy2014},
    for $\text{C}_{2}\text{H}_{6}$ was taken from Ref.\citenum{liDrivenSimilarityRenormalization2017}.
  \end{tablenotes}
\end{threeparttable} \endgroup
\end{table}

\begin{figure*}
\centering \begin{subfigure}[b]{0.475\textwidth} \centering
\includegraphics[width=1\textwidth]{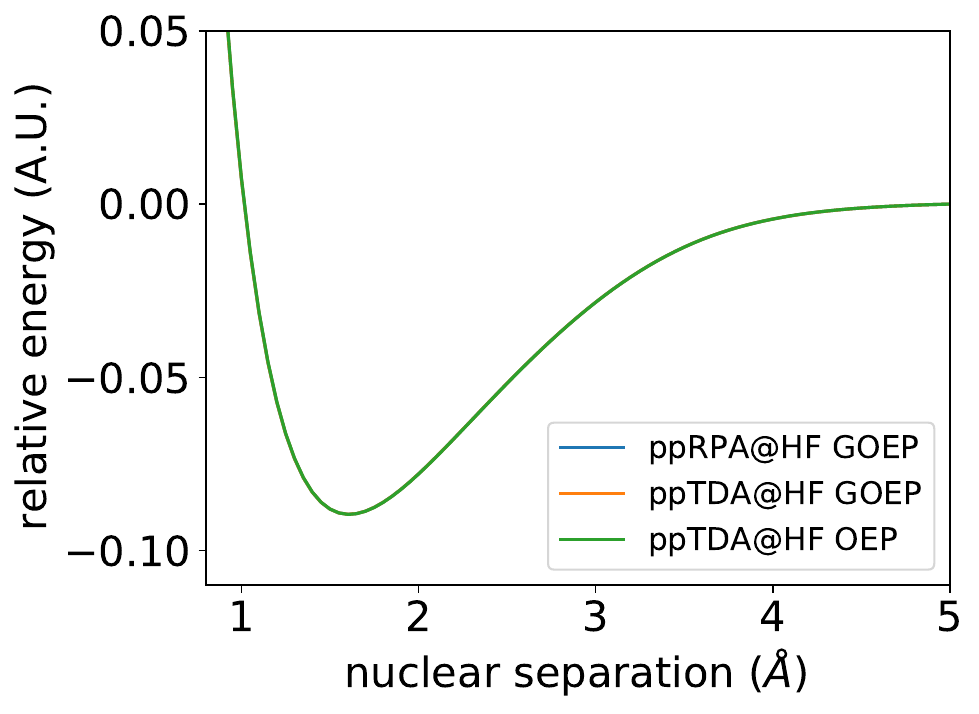} \caption{LiH}
\label{fig:geop_lih} \end{subfigure} \hfill{}\begin{subfigure}[b]{0.475\textwidth}
\centering \includegraphics[width=1\textwidth]{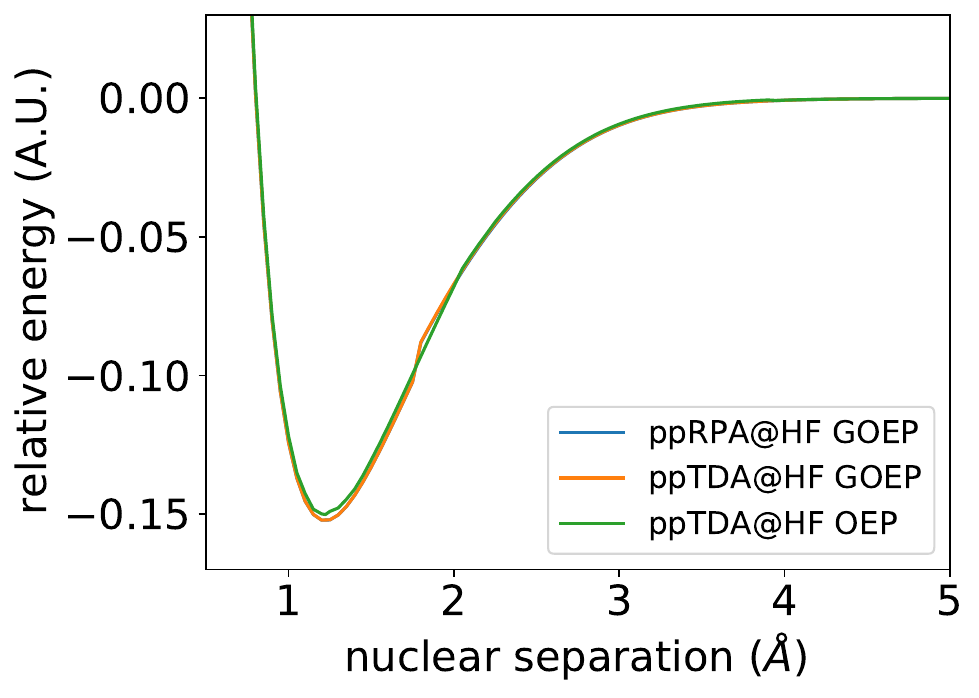}
\caption{BH}
\label{fig:goep_bh} \end{subfigure} \caption{Dissociation curves of LiH and BH obtained from ppTDA@HF optimized
with the OEP method, ppTDA@HF optimized with the GOEP method, and
ppRPA@HF optimized with the GOEP method. The cc-pVTZ basis set was
used.}
\label{fig:goep}
\end{figure*}

% Results: OEP vs GOEP
We now show that the SCF solution of the multireference DFT approach
is not affected by the choice of the optimization methods with OEP or GOEP,
and the choice of methods for the two-electron addition energy from ppTDA or ppRPA.
Dissociation curves of LiH and BH obtained from ppTDA@HF
optimized with the OEP method, ppTDA@HF optimized with the GOEP method
\cite{chenMultireferenceDensityFunctional2017} and ppRPA@HF optimized
with the GEOP method are shown in Fig.\ref{fig:goep}. Comparing dissociation
curves of ppTDA@HF optimized with OEP and GEOP, two optimization methods
produce close results for both LiH and BH. We used the OEP method
in the rest of this work because it produces physical potentials for most systems
(with the exception of Be atom, which is discussed in the Section.S2
of the Supporting Information), while the GOEP method can produce
unphysical potentials with a negative HOMO-LUMO gap when combining
with traditional DFAs.
Comparing dissociation curves in Fig.\ref{fig:goep} of ppTDA@HF and
ppRPA@HF methods optimized with GOEP, we see that using ppRPA and ppTDA for
excitation energies lead to very close results because they have similar
physics for two-electron addition energies\cite{yangDoubleRydbergCharge2013,yangExcitationEnergiesParticleparticle2014}.
We used ppTDA in the rest of this work for simplicity. \\

\begin{figure}
\centering \includegraphics[width=1\textwidth]{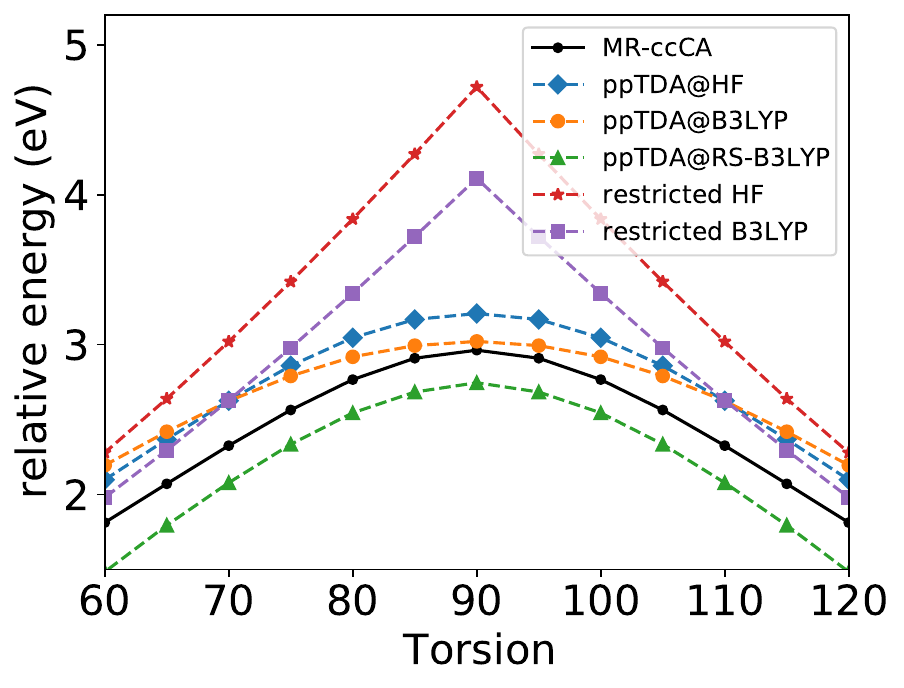}
\caption{Relative energy curves of twisted ethylene obtained from ppTDA@HF,
ppTDA@B3LYP, ppTDA@RS-B3LYP, restricted HF and restricted B3LYP at
$\theta=$\ang{0}, \ang{60}, \ang{65}, \ang{70}, \ang{75},
\ang{80}, \ang{85} and \ang{90}. The energy at $\theta=0$
is set as zero. The MR-ccCA results taken from Ref.\citenum{jiangEmpiricalCorrectionNondynamical2012}
was used as the reference. All curves are shifted such that the energy
at \ang{0} is zero. Structures were taken from Ref.\citenum{chenMultireferenceDensityFunctional2017}.
The cc-pVTZ basis set was used.}
\label{fig:torsion}
\end{figure}

% Results: twisted ethylene
Another example of the ppTDA@RS-DFA approach successfully describing
the static correlation is the ethylene torsion. Relative energy curves
along the ethylene torsional coordinate obtained from ppTDA@HF, ppTDA@B3LYP,
ppTDA@RS-B3LYP, restricted HF and restricted B3LYP are shown in Fig.\ref{fig:torsion}.
The MR-ccCA results from Ref.\citenum{jiangEmpiricalCorrectionNondynamical2012}
were used as the reference. Because of the diradical character, the
static correlation is important in the twisted ethylene. Restricted
HF and restricted B3LYP that have poor descriptions for the static
correlation produce unphysical relative energy curves and large errors
in the barrier heights. It has been shown that unrestricted KS-DFT
can provide smooth curves for the ethylene rotation\cite{shaoSpinFlipApproach2003}.
However, finding an appropriate spin-symmetry broken solution is difficult
in practical calculations\cite{shaoSpinFlipApproach2003}. With the
proper multiconfigurational description, ppTDA@HF, ppTDA@B3LYP and
ppTDA@RS-B3LYP provide qualitatively correct relative energies at
all angles. Although ppTDA@B3LYP gives an accurate barrier height,
it shows increasing deviations at both ends. ppTDA@HF and ppTDA@RS-B3LYP
provide accurate curves that are in good agreement with the curve
of MR-ccCA for all angles. \\

% Results: excitation energies
In addition to proper descriptions of ground state properties, our
ppTDA@RS-DFA approach also predicts accurate excitation energies.
Excitation energies were obtained by the energy difference between
the target excited state and the ground state. To demonstrate the
accuracy, excitation energies of molecular and atomic systems obtained
from ppTDA@DFA, and ppTDA@RS-DFA, and post-SCF approaches post-SCF-ppTDA@DFA
and post-SCF-ppTDA@RS-DFA as well as traditional TDDFT are listed in Table.\ref{tab:excitation}. %We first find results of ppTDA@HF are close to results of ppRPA@HF reported in
%Ref.\citenum{chenMultireferenceDensityFunctional2017},
%which agrees with the observation for dissociation curves.
It shows that the conventional post-SCF approach post-SCF-ppTDA@DFA provides
an adequate accuracy. B3LYP is the best starting point for post-SCF-ppTDA@DFA
with a MAE of $0.48$ \,{eV}. Only introducing the self-consistency
or applying RS does not improve the accuracy. MAEs of post-SCF-ppTDA@RS-DFA
are around $1.0$ \,{eV} and MAEs of ppTDA@DFA can even exceed
$2.0$ \,{eV}. Combining the self-consistency and RS leads to considerable
improvement. ppTDA@RS-B3LYP and ppTDA@RS-PBE provide the smallest
MAEs of only $0.32$ \,{eV} that are similar to ppTDA@HF. Besides
the improved accuracy, the dependence on the choice of the DFA is
also largely eliminated in the ppTDA@RS-DFA approach. ppTDA@RS-B3LYP
and ppTDA@RS-PBE provide very similar results for all tested systems.
ppTDA@RS-DFA also outperforms the traditional TDDFT approach.
As shown in Table.\ref{tab:excitation},
TDDFT gives large errors for tested systems.
TDDFT even provides negative excitation energies for triplet excitation BH,
CH$^+$ and Ca,
which are related to the triplet instability\cite{peachTripletInstabilityTDDFT2013,casidaProgressTimeDependentDensityFunctional2012}.
Our ppTDA@RS-DFA approach describes excited states of tested small molecules and
atoms well.
The performance for larger systems needs to be further investigated.

\begin{table*}
        \small
        \setlength\tabcolsep{3.5pt}
\caption{Excitation energies of molecular and atomic systems obtained from
ppTDA@DFA, ppTDA@RS-DFA, post-SCF-ppTDA@DFA, post-SCF-ppTDA@RS-DFA and TDDFT@DFA,
all values in {eV}.}
\label{tab:excitation} \begin{threeparttable} %
\begin{tabular*}{0.99\linewidth}{p{0.8cm}cccccccccccccccc}
\hline
 &  &  & \multicolumn{5}{c}{ppTDA} & \multicolumn{5}{c}{post-SCF-ppTDA} & \multicolumn{3}{c}{TD}\tabularnewline
\hline
 &  & ref\tnote{2}  & a\tnote{3}  & b\tnote{3}  & c\tnote{3}  & d\tnote{3}  & e\tnote{3}  & a  & b  & c  & d  & e & a  & b & d\tabularnewline
\hline
BH                       & Triplet  & 1.27  & 1.54  & 1.72  & 1.46  & 1.89  & 1.46  & 1.63  & 1.33  & 1.66  & 1.19  & 1.67 & -1.63 & -0.44 &       \tabularnewline
                         & Singlet  & 2.85  & 3.45  & 3.53  & 3.25  & 3.63  & 3.21  & 3.15  & 3.11  & 3.14  & 3.01  & 3.13 & 2.66  & 2.69  &       \tabularnewline
CH$^{+}$                 & Triplet  & 1.15  & 1.48  & 0.68  & 1.36  & 0.62  & 1.35  & 1.66  & 1.06  & 1.69  & 0.91  & 1.69 & -1.86 & -1.23 &       \tabularnewline
                         & Singlet  & 3.07  & 3.81  & 2.87  & 3.55  & 2.79  & 3.50  & 3.54  & 3.10  & 3.53  & 2.98  & 3.52 & 2.65  & 2.83  &       \tabularnewline
LiH\tnote{1}             & Triplet  & 3.25  & 3.23  &       & 3.23  &       & 3.23  & 3.16  & 3.53  & 3.17  & 3.49  & 3.17 & 2.80  & 2.65  & 2.35  \tabularnewline
                         & Singlet  & 3.61  & 3.60  &       & 3.60  &       & 3.60  & 3.53  & 3.89  & 3.53  & 3.85  & 3.53 & 4.04  & 3.21  & 2.99  \tabularnewline
CO                       & Triplet  & 6.32  & 6.45  & 6.07  & 7.07  & 10.23  & 7.24  & 5.61  & 5.82  & 6.18  & 5.80  & 6.35& 5.28  & 5.86  & 5.74  \tabularnewline
                         & Singlet  & 8.51  & 9.35  & 8.67  & 9.37  & 12.65  & 9.42  & 7.86  & 8.15  & 8.26  & 8.13  & 8.40& 8.79  & 8.40  & 8.25  \tabularnewline
C$_{2}$H$_{4}$\tnote{1}  & Triplet  & 4.50  & 4.50  & 3.83  & 4.50  & 3.64  & 4.49  & 3.95  & 3.64  & 3.92  & 3.49  & 3.91 & 1.05  & 4.12  & 4.31  \tabularnewline
                         & Singlet  & 7.80  & 6.93  & 10.15  & 6.78  & 10.39  & 6.77  & 4.54  & 7.82  & 4.45  & 8.63  & 4.45 & 7.16 & 6.65 & 6.51 \tabularnewline
$\text{H}_{2}\text{O}$   & Triplet  & 7.00  & 6.71  & 10.94  & 6.70  & 12.02  & 6.78  & 5.98  & 6.22  & 3.08  & 7.14  & 3.14 & 7.79 & 6.48 & 6.00\tabularnewline
                         & Singlet  & 7.40  & 7.05  & 11.75  & 7.03  & 13.00  & 7.11  & 5055  & 6.63  & 3.31  & 7.62  & 3.36 & 8.57 & 6.86 & 6.35 \tabularnewline
Mg                       & Triplet  & 2.71  & 4.28  &  & 2.61  &  & 2.61  & 2.59  & 3.45  & 2.60  & 3.53  & 2.60 & 1.11 & 2.61 & 2.33 \tabularnewline
                         & Singlet  & 4.35  & 1.69  &  & 4.30  &  & 4.30  & 4.27  & 5.69  & 4.28  & 5.76  & 4.29 & 4.06 & 4.23 & 4.18 \tabularnewline
Ca\tnote{1}              & Triplet  & 1.79  & 1.69  &  & 1.70  &  & 1.70  & 1.68  & 0.96  & 1.68  & 0.13  & 1.68 & -0.48 & 1.52 & 1.16\tabularnewline
\hline
 &  & MAE  & 0.32  & 1.21  & 0.32  & 2.06  & 0.32  & 0.60  & 0.48  & 0.96  & 0.54  & 0.94 & 1.26 & 0.61 & 0.64\tabularnewline
\hline
\end{tabular*}\begin{tablenotes}

\item{[}1{]} The cc-pVTZ basis set was used for LiH and Ca. The aug-cc-pVDZ
basis set was used for $\text{C}_{2}\text{H}_{4}$. The aug-cc-pVTZ
was used for remaining systems.

\item{[}2{]} The reference value for LiH was obtained from the EOM-CCSD
calculation. The reference value for $\text{H}_{2}\text{O}$ was taken
from Ref.\citenum{hoyerMulticonfigurationPairDensityFunctional2016}.
Reference values for remaining systems were taken from Ref.\citenum{chenMultireferenceDensityFunctional2017}.

\item{[}3{]} a=HF, b=B3LYP, c=RS-B3LYP, d=PBE, e=RS-PBE
\end{tablenotes} \end{threeparttable}
\end{table*}

% scaling discussion
The ppTDA@RS-DFA approach has a favorable computational scaling. For
the energy, the scaling of evaluating the energy of the ($N-2$)-electron
system is $\mathcal{O}(N^{3})$ for a LDA or GGA functional or $\mathcal{O}(N^{4})$
for a hybrid functional, where $N$ is the size of the system. The
scaling for evaluating the excitation energy from the ppTDA equation
in Eq.\ref{eq:pptda} is $\mathcal{O}(N^{4})$ with the Davidson algorithm\cite{yangExcitationEnergiesParticleparticle2014}.
For the gradient, the scaling of evaluating the gradient of the ($N-2$)-electron
system energy is $\mathcal{O}(N^{3})$. The evaluation of the gradient
of the excitation energy is the bottleneck. For ppTDA@DFA where the
DFA is not HF, the evaluation of the last term in Eq.\ref{eq:gradient_pp}
is the dominant step. Because the Hxc kernel used in Eq.\ref{eq:gradient_pp}
is a four-index quantity that cannot be decomposed into a summation
with three-index matrices by the RI approximation, this leads the
$\mathcal{O}(N^{5})$ scaling for evaluating Eq.\ref{eq:gradient_pp}.
However, for ppTDA@RS-DFA, the Hxc kernel only involves two-electron integrals
as shown in Eq.\ref{eq:rs_kernel}. Therefore, the last term in Eq.\ref{eq:gradient_pp}
can be evaluated of $\mathcal{O}(N^{4})$ as
\begin{equation}
X_{a'b'}^{n}K_{a'c',ai}^{\text{RS,Hxc}}X_{c'b'}^{n}=\sum_{P}\bigg[X_{a'b'}^{n}\bigg(R_{P,a'c'}R_{P,ai}-R_{P,c'a}R_{P,a'i}\bigg)X_{c'b'}^{n}\bigg]\text{,}\label{eq:gradient_ri}
\end{equation}
where the intermediate quantity $R_{P,pq}=\sum_{Q}(P|Q)^{-\frac{1}{2}}(Q|rs)$
is constructed in the RI approximation and $\{\tilde{\psi}_{Q}\}$
is a set of auxiliary basis sets. By using Eq.\ref{eq:gradient_ri},
the scaling of evaluating the gradient of excitation energy part is
$\mathcal{O}(N^{4})$. Thus the formal scaling of the ppTDA@RS-DFA
approach is $\mathcal{O}(N^{4})$, which is the same as the SCF HF/hybrid
functional calculation. The computational cost of ppTDA@RS-DFA can
be further reduced significantly by using the active space method\cite{zhangAccurateEfficientCalculation2016}.
In the active space method, the excitation energy is obtained by a
direct diagonalization of the ppTDA equation in a small subspace.
The scaling of solving the ppTDA equation with the active space method
is $\mathcal{O}(N^{4})$, with a prefactor much smaller than a single
SCF HF/hybrid functional calculation\cite{zhangAccurateEfficientCalculation2016}.
Because of using the active space, the cost of evaluating the gradient
is also greatly reduced because the two-electron addition eigenvector
has a smaller dimension. Therefore, the ppTDA@RS-DFA approach is promising
for application to larger systems.

% Conclusion
In summary, we applied the RS approach in the multireference DFT approach
to describe ground states and excited states of systems with static correlation.
In ppTDA@RS-DFA,
the RS Hamiltonian is used in the ppTDA equation to include all contributions
from one-body perturbations. Then the SCF solution of the total energy
is obtained from the optimization of the OEP method of Yang and Wu.
We showed that the OEP method produces physical potential for most
systems, which are not guaranteed in GOEP optimizations when using
traditional DFAs. The ppTDA@RS-DFA approach was first examined on
describing single bond breaking problems. We showed that ppTDA@RS-DFA
provides accurate bond dissociation curves with proper spin density symmetry,
dissociation energies and equilibrium bond lengths for all four systems,
significantly outperforming ppTDA@DFA and conventional KS-DFT.
Then we showed that the ppTDA@RS-DFA approach also successfully describes
the double bond rotation in ethylene.
We also demonstrated that the ppTDA@RS-DFA approach is capable of
describing excited states.
ppTDA@RS-DFA provides accurate excitation
energies for both molecular and atomic systems. And the dependence
on the choice of the DFA in multireference DFT is largely eliminated.
Beyond the ground state and excited state properties,
for state-to-state properties such as the conical intersection search and couplings,
further study is needed.
Our ppTDA@RS-DFA approach scales as $\mathcal{O}(N^{4})$. The computational
cost can be further reduced by using the active space method.
The ppTDA@RS-DFA approach is expected to open up new possibilities for the
application of the multireference DFT approach for large systems.

% Acknowledgement
\begin{acknowledgement}
ACKNOWLEDGMENTS: J. L. and Z.C. acknowledge the support from the National
Institute of General Medical Sciences of the National Institutes of
Health under award number R01-GM061870. W.Y. acknowledges the support
from the National Science Foundation (grant no. CHE-1900338).
\end{acknowledgement}
\begin{suppinfo}
Supporting Information Available: orbital optimization for the total energy in multireference DFT, examples of unphysical energy in the optimization, comparison of choices of the fixed density, discussion of degeneracy in multireference DFT and potential functional theory.
\end{suppinfo}

\bibliography{ref,software}

\providecommand{\latin}[1]{#1}
\makeatletter
\providecommand{\doi}
  {\begingroup\let\do\@makeother\dospecials
  \catcode`\{=1 \catcode`\}=2 \doi@aux}
\providecommand{\doi@aux}[1]{\endgroup\texttt{#1}}
\makeatother
\providecommand*\mcitethebibliography{\thebibliography}
\csname @ifundefined\endcsname{endmcitethebibliography}
  {\let\endmcitethebibliography\endthebibliography}{}
\begin{mcitethebibliography}{80}
\providecommand*\natexlab[1]{#1}
\providecommand*\mciteSetBstSublistMode[1]{}
\providecommand*\mciteSetBstMaxWidthForm[2]{}
\providecommand*\mciteBstWouldAddEndPuncttrue
  {\def\EndOfBibitem{\unskip.}}
\providecommand*\mciteBstWouldAddEndPunctfalse
  {\let\EndOfBibitem\relax}
\providecommand*\mciteSetBstMidEndSepPunct[3]{}
\providecommand*\mciteSetBstSublistLabelBeginEnd[3]{}
\providecommand*\EndOfBibitem{}
\mciteSetBstSublistMode{f}
\mciteSetBstMaxWidthForm{subitem}{(\alph{mcitesubitemcount})}
\mciteSetBstSublistLabelBeginEnd
  {\mcitemaxwidthsubitemform\space}
  {\relax}
  {\relax}

\bibitem[Hohenberg and Kohn(1964)Hohenberg, and
  Kohn]{hohenbergInhomogeneousElectronGas1964}
Hohenberg,~P.; Kohn,~W. Inhomogeneous {{Electron Gas}}. \emph{Phys. Rev.}
  \textbf{1964}, \emph{136}, B864--B871\relax
\mciteBstWouldAddEndPuncttrue
\mciteSetBstMidEndSepPunct{\mcitedefaultmidpunct}
{\mcitedefaultendpunct}{\mcitedefaultseppunct}\relax
\EndOfBibitem
\bibitem[Kohn and Sham(1965)Kohn, and
  Sham]{kohnSelfConsistentEquationsIncluding1965}
Kohn,~W.; Sham,~L.~J. Self-{{Consistent Equations Including Exchange}} and
  {{Correlation Effects}}. \emph{Phys. Rev.} \textbf{1965}, \emph{140},
  A1133--A1138\relax
\mciteBstWouldAddEndPuncttrue
\mciteSetBstMidEndSepPunct{\mcitedefaultmidpunct}
{\mcitedefaultendpunct}{\mcitedefaultseppunct}\relax
\EndOfBibitem
\bibitem[Parr and Weitao(1989)Parr, and
  Weitao]{parrDensityFunctionalTheoryAtoms1989}
Parr,~R.~G.; Weitao,~Y. \emph{Density-{{Functional Theory}} of {{Atoms}} and
  {{Molecules}}}; {Oxford University Press}, 1989\relax
\mciteBstWouldAddEndPuncttrue
\mciteSetBstMidEndSepPunct{\mcitedefaultmidpunct}
{\mcitedefaultendpunct}{\mcitedefaultseppunct}\relax
\EndOfBibitem
\bibitem[von Barth and Hedin(1972)von Barth, and
  Hedin]{barthLocalExchangecorrelationPotential1972}
von Barth,~U.; Hedin,~L. A Local Exchange-Correlation Potential for the Spin
  Polarized Case. i. \emph{J. Phys. C: Solid State Phys.} \textbf{1972},
  \emph{5}, 1629--1642\relax
\mciteBstWouldAddEndPuncttrue
\mciteSetBstMidEndSepPunct{\mcitedefaultmidpunct}
{\mcitedefaultendpunct}{\mcitedefaultseppunct}\relax
\EndOfBibitem
\bibitem[Vosko \latin{et~al.}(1980)Vosko, Wilk, and
  Nusair]{voskoAccurateSpindependentElectron1980}
Vosko,~S.~H.; Wilk,~L.; Nusair,~M. Accurate Spin-Dependent Electron Liquid
  Correlation Energies for Local Spin Density Calculations: A Critical
  Analysis. \emph{Can. J. Phys.} \textbf{1980}, \emph{58}, 1200--1211\relax
\mciteBstWouldAddEndPuncttrue
\mciteSetBstMidEndSepPunct{\mcitedefaultmidpunct}
{\mcitedefaultendpunct}{\mcitedefaultseppunct}\relax
\EndOfBibitem
\bibitem[Becke(1988)]{beckeDensityfunctionalExchangeenergyApproximation1988}
Becke,~A.~D. Density-Functional Exchange-Energy Approximation with Correct
  Asymptotic Behavior. \emph{Phys. Rev. A} \textbf{1988}, \emph{38},
  3098--3100\relax
\mciteBstWouldAddEndPuncttrue
\mciteSetBstMidEndSepPunct{\mcitedefaultmidpunct}
{\mcitedefaultendpunct}{\mcitedefaultseppunct}\relax
\EndOfBibitem
\bibitem[Lee \latin{et~al.}(1988)Lee, Yang, and
  Parr]{leeDevelopmentColleSalvettiCorrelationenergy1988}
Lee,~C.; Yang,~W.; Parr,~R.~G. Development of the {{Colle-Salvetti}}
  Correlation-Energy Formula into a Functional of the Electron Density.
  \emph{Phys. Rev. B} \textbf{1988}, \emph{37}, 785--789\relax
\mciteBstWouldAddEndPuncttrue
\mciteSetBstMidEndSepPunct{\mcitedefaultmidpunct}
{\mcitedefaultendpunct}{\mcitedefaultseppunct}\relax
\EndOfBibitem
\bibitem[Perdew and Wang(1992)Perdew, and
  Wang]{perdewAccurateSimpleAnalytic1992}
Perdew,~J.~P.; Wang,~Y. Accurate and Simple Analytic Representation of the
  Electron-Gas Correlation Energy. \emph{Phys. Rev. B} \textbf{1992},
  \emph{45}, 13244--13249\relax
\mciteBstWouldAddEndPuncttrue
\mciteSetBstMidEndSepPunct{\mcitedefaultmidpunct}
{\mcitedefaultendpunct}{\mcitedefaultseppunct}\relax
\EndOfBibitem
\bibitem[Perdew \latin{et~al.}(1996)Perdew, Burke, and
  Ernzerhof]{perdewGeneralizedGradientApproximation1996}
Perdew,~J.~P.; Burke,~K.; Ernzerhof,~M. Generalized {{Gradient Approximation
  Made Simple}}. \emph{Phys. Rev. Lett.} \textbf{1996}, \emph{77},
  3865--3868\relax
\mciteBstWouldAddEndPuncttrue
\mciteSetBstMidEndSepPunct{\mcitedefaultmidpunct}
{\mcitedefaultendpunct}{\mcitedefaultseppunct}\relax
\EndOfBibitem
\bibitem[Becke(1993)]{beckeNewMixingHartree1993}
Becke,~A.~D. A New Mixing of {{Hartree}}\textendash{{Fock}} and Local
  Density-functional Theories. \emph{J. Chem. Phys.} \textbf{1993}, \emph{98},
  1372--1377\relax
\mciteBstWouldAddEndPuncttrue
\mciteSetBstMidEndSepPunct{\mcitedefaultmidpunct}
{\mcitedefaultendpunct}{\mcitedefaultseppunct}\relax
\EndOfBibitem
\bibitem[Zhang \latin{et~al.}(2021)Zhang, Hermes, Gagliardi, and
  Truhlar]{zhangMulticonfigurationDensityCoherenceFunctional2021}
Zhang,~D.; Hermes,~M.~R.; Gagliardi,~L.; Truhlar,~D.~G. Multiconfiguration
  {{Density-Coherence Functional Theory}}. \emph{J. Chem. Theory Comput.}
  \textbf{2021}, \emph{17}, 2775--2782\relax
\mciteBstWouldAddEndPuncttrue
\mciteSetBstMidEndSepPunct{\mcitedefaultmidpunct}
{\mcitedefaultendpunct}{\mcitedefaultseppunct}\relax
\EndOfBibitem
\bibitem[Li~Manni \latin{et~al.}(2014)Li~Manni, Carlson, Luo, Ma, Olsen,
  Truhlar, and Gagliardi]{limanniMulticonfigurationPairDensityFunctional2014}
Li~Manni,~G.; Carlson,~R.~K.; Luo,~S.; Ma,~D.; Olsen,~J.; Truhlar,~D.~G.;
  Gagliardi,~L. Multiconfiguration {{Pair-Density Functional Theory}}. \emph{J.
  Chem. Theory Comput.} \textbf{2014}, \emph{10}, 3669--3680\relax
\mciteBstWouldAddEndPuncttrue
\mciteSetBstMidEndSepPunct{\mcitedefaultmidpunct}
{\mcitedefaultendpunct}{\mcitedefaultseppunct}\relax
\EndOfBibitem
\bibitem[Ren \latin{et~al.}(2013)Ren, Rinke, Scuseria, and
  Scheffler]{renRenormalizedSecondorderPerturbation2013}
Ren,~X.; Rinke,~P.; Scuseria,~G.~E.; Scheffler,~M. Renormalized Second-Order
  Perturbation Theory for the Electron Correlation Energy: {{Concept}},
  Implementation, and Benchmarks. \emph{Phys. Rev. B} \textbf{2013}, \emph{88},
  035120\relax
\mciteBstWouldAddEndPuncttrue
\mciteSetBstMidEndSepPunct{\mcitedefaultmidpunct}
{\mcitedefaultendpunct}{\mcitedefaultseppunct}\relax
\EndOfBibitem
\bibitem[Cohen \latin{et~al.}(2008)Cohen, {Mori-S{\'a}nchez}, and
  Yang]{cohenInsightsCurrentLimitations2008}
Cohen,~A.~J.; {Mori-S{\'a}nchez},~P.; Yang,~W. Insights into {{Current
  Limitations}} of {{Density Functional Theory}}. \emph{Science} \textbf{2008},
  \emph{321}, 792--794\relax
\mciteBstWouldAddEndPuncttrue
\mciteSetBstMidEndSepPunct{\mcitedefaultmidpunct}
{\mcitedefaultendpunct}{\mcitedefaultseppunct}\relax
\EndOfBibitem
\bibitem[Cohen \latin{et~al.}(2008)Cohen, {Mori-S{\'a}nchez}, and
  Yang]{cohenFractionalSpinsStatic2008}
Cohen,~A.~J.; {Mori-S{\'a}nchez},~P.; Yang,~W. Fractional Spins and Static
  Correlation Error in Density Functional Theory. \emph{J. Chem. Phys.}
  \textbf{2008}, \emph{129}, 121104\relax
\mciteBstWouldAddEndPuncttrue
\mciteSetBstMidEndSepPunct{\mcitedefaultmidpunct}
{\mcitedefaultendpunct}{\mcitedefaultseppunct}\relax
\EndOfBibitem
\bibitem[Roos \latin{et~al.}(1980)Roos, Taylor, and
  Sigbahn]{roosCompleteActiveSpace1980}
Roos,~B.~O.; Taylor,~P.~R.; Sigbahn,~P. E.~M. A Complete Active Space {{SCF}}
  Method ({{CASSCF}}) Using a Density Matrix Formulated Super-{{CI}} Approach.
  \emph{Chem. Phys.} \textbf{1980}, \emph{48}, 157--173\relax
\mciteBstWouldAddEndPuncttrue
\mciteSetBstMidEndSepPunct{\mcitedefaultmidpunct}
{\mcitedefaultendpunct}{\mcitedefaultseppunct}\relax
\EndOfBibitem
\bibitem[Roos(1983)]{roosMulticonfigurationalMCSCF1983}
Roos,~B.~O. In \emph{Methods in {{Computational Molecular Physics}}};
  Diercksen,~G. H.~F., Wilson,~S., Eds.; {{NATO ASI Series}}; {Springer
  Netherlands}: {Dordrecht}, 1983; pp 161--187\relax
\mciteBstWouldAddEndPuncttrue
\mciteSetBstMidEndSepPunct{\mcitedefaultmidpunct}
{\mcitedefaultendpunct}{\mcitedefaultseppunct}\relax
\EndOfBibitem
\bibitem[Gagliardi \latin{et~al.}(2017)Gagliardi, Truhlar, Li~Manni, Carlson,
  Hoyer, and Bao]{gagliardiMulticonfigurationPairDensityFunctional2017}
Gagliardi,~L.; Truhlar,~D.~G.; Li~Manni,~G.; Carlson,~R.~K.; Hoyer,~C.~E.;
  Bao,~J.~L. Multiconfiguration {{Pair-Density Functional Theory}}: {{A New Way
  To Treat Strongly Correlated Systems}}. \emph{Acc. Chem. Res.} \textbf{2017},
  \emph{50}, 66--73\relax
\mciteBstWouldAddEndPuncttrue
\mciteSetBstMidEndSepPunct{\mcitedefaultmidpunct}
{\mcitedefaultendpunct}{\mcitedefaultseppunct}\relax
\EndOfBibitem
\bibitem[Ghosh \latin{et~al.}(2015)Ghosh, Sonnenberger, Hoyer, Truhlar, and
  Gagliardi]{ghoshMulticonfigurationPairDensityFunctional2015}
Ghosh,~S.; Sonnenberger,~A.~L.; Hoyer,~C.~E.; Truhlar,~D.~G.; Gagliardi,~L.
  Multiconfiguration {{Pair-Density Functional Theory Outperforms
  Kohn}}\textendash{{Sham Density Functional Theory}} and {{Multireference
  Perturbation Theory}} for {{Ground-State}} and {{Excited-State Charge
  Transfer}}. \emph{J. Chem. Theory Comput.} \textbf{2015}, \emph{11},
  3643--3649\relax
\mciteBstWouldAddEndPuncttrue
\mciteSetBstMidEndSepPunct{\mcitedefaultmidpunct}
{\mcitedefaultendpunct}{\mcitedefaultseppunct}\relax
\EndOfBibitem
\bibitem[Hoyer \latin{et~al.}(2016)Hoyer, Ghosh, Truhlar, and
  Gagliardi]{hoyerMulticonfigurationPairDensityFunctional2016}
Hoyer,~C.~E.; Ghosh,~S.; Truhlar,~D.~G.; Gagliardi,~L. Multiconfiguration
  {{Pair-Density Functional Theory Is}} as {{Accurate}} as {{CASPT2}} for
  {{Electronic Excitation}}. \emph{J. Phys. Chem. Lett.} \textbf{2016},
  \emph{7}, 586--591\relax
\mciteBstWouldAddEndPuncttrue
\mciteSetBstMidEndSepPunct{\mcitedefaultmidpunct}
{\mcitedefaultendpunct}{\mcitedefaultseppunct}\relax
\EndOfBibitem
\bibitem[Grofe \latin{et~al.}(2017)Grofe, Chen, Liu, and
  Gao]{grofeSpinMultipletComponentsEnergy2017}
Grofe,~A.; Chen,~X.; Liu,~W.; Gao,~J. Spin-{{Multiplet Components}} and
  {{Energy Splittings}} by {{Multistate Density Functional Theory}}. \emph{J.
  Phys. Chem. Lett.} \textbf{2017}, \emph{8}, 4838--4845\relax
\mciteBstWouldAddEndPuncttrue
\mciteSetBstMidEndSepPunct{\mcitedefaultmidpunct}
{\mcitedefaultendpunct}{\mcitedefaultseppunct}\relax
\EndOfBibitem
\bibitem[Ren \latin{et~al.}(2016)Ren, Provorse, Bao, Qu, and
  Gao]{renMultistateDensityFunctional2016}
Ren,~H.; Provorse,~M.~R.; Bao,~P.; Qu,~Z.; Gao,~J. Multistate {{Density
  Functional Theory}} for {{Effective Diabatic Electronic Coupling}}. \emph{J.
  Phys. Chem. Lett.} \textbf{2016}, \emph{7}, 2286--2293\relax
\mciteBstWouldAddEndPuncttrue
\mciteSetBstMidEndSepPunct{\mcitedefaultmidpunct}
{\mcitedefaultendpunct}{\mcitedefaultseppunct}\relax
\EndOfBibitem
\bibitem[Gao \latin{et~al.}(2016)Gao, Grofe, Ren, and
  Bao]{gaoKohnShamApproximation2016}
Gao,~J.; Grofe,~A.; Ren,~H.; Bao,~P. Beyond {{Kohn}}\textendash{{Sham
  Approximation}}: {{Hybrid Multistate Wave Function}} and {{Density Functional
  Theory}}. \emph{J. Phys. Chem. Lett.} \textbf{2016}, \emph{7},
  5143--5149\relax
\mciteBstWouldAddEndPuncttrue
\mciteSetBstMidEndSepPunct{\mcitedefaultmidpunct}
{\mcitedefaultendpunct}{\mcitedefaultseppunct}\relax
\EndOfBibitem
\bibitem[Su \latin{et~al.}(2018)Su, Li, and
  Yang]{suDescribingStrongCorrelation2018}
Su,~N.~Q.; Li,~C.; Yang,~W. Describing Strong Correlation with Fractional-Spin
  Correction in Density Functional Theory. \emph{PNAS} \textbf{2018},
  \emph{115}, 9678--9683\relax
\mciteBstWouldAddEndPuncttrue
\mciteSetBstMidEndSepPunct{\mcitedefaultmidpunct}
{\mcitedefaultendpunct}{\mcitedefaultseppunct}\relax
\EndOfBibitem
\bibitem[Filatov(2015)]{filatovSpinrestrictedEnsemblereferencedKohn2015}
Filatov,~M. Spin-Restricted Ensemble-Referenced {{Kohn}}\textendash{{Sham}}
  Method: Basic Principles and Application to Strongly Correlated Ground and
  Excited States of Molecules. \emph{WIREs Comput. Mol. Sci.} \textbf{2015},
  \emph{5}, 146--167\relax
\mciteBstWouldAddEndPuncttrue
\mciteSetBstMidEndSepPunct{\mcitedefaultmidpunct}
{\mcitedefaultendpunct}{\mcitedefaultseppunct}\relax
\EndOfBibitem
\bibitem[Ullrich(2011)]{ullrichTimeDependentDensityFunctionalTheory2011}
Ullrich,~C.~A. \emph{Time-{{Dependent Density-Functional Theory}}: {{Concepts}}
  and {{Applications}}}; {OUP Oxford}, 2011\relax
\mciteBstWouldAddEndPuncttrue
\mciteSetBstMidEndSepPunct{\mcitedefaultmidpunct}
{\mcitedefaultendpunct}{\mcitedefaultseppunct}\relax
\EndOfBibitem
\bibitem[Runge and Gross(1984)Runge, and
  Gross]{rungeDensityFunctionalTheoryTimeDependent1984}
Runge,~E.; Gross,~E. K.~U. Density-{{Functional Theory}} for {{Time-Dependent
  Systems}}. \emph{Phys. Rev. Lett.} \textbf{1984}, \emph{52}, 997--1000\relax
\mciteBstWouldAddEndPuncttrue
\mciteSetBstMidEndSepPunct{\mcitedefaultmidpunct}
{\mcitedefaultendpunct}{\mcitedefaultseppunct}\relax
\EndOfBibitem
\bibitem[Casida(1995)]{casidaTimeDependentDensityFunctional1995}
Casida,~M.~E. \emph{Recent {{Advances}} in {{Density Functional Methods}}};
  Recent {{Advances}} in {{Computational Chemistry}}; {WORLD SCIENTIFIC}, 1995;
  Vol. Volume 1; pp 155--192\relax
\mciteBstWouldAddEndPuncttrue
\mciteSetBstMidEndSepPunct{\mcitedefaultmidpunct}
{\mcitedefaultendpunct}{\mcitedefaultseppunct}\relax
\EndOfBibitem
\bibitem[Casida and {Huix-Rotllant}(2012)Casida, and
  {Huix-Rotllant}]{casidaProgressTimeDependentDensityFunctional2012}
Casida,~M.; {Huix-Rotllant},~M. Progress in {{Time-Dependent Density-Functional
  Theory}}. \emph{Annu. Rev. Phys. Chem.} \textbf{2012}, \emph{63},
  287--323\relax
\mciteBstWouldAddEndPuncttrue
\mciteSetBstMidEndSepPunct{\mcitedefaultmidpunct}
{\mcitedefaultendpunct}{\mcitedefaultseppunct}\relax
\EndOfBibitem
\bibitem[Peach \latin{et~al.}(2013)Peach, Warner, and
  Tozer]{peachTripletInstabilityTDDFT2013}
Peach,~M.~J.; Warner,~N.; Tozer,~D.~J. On the Triplet Instability in {{TDDFT}}.
  \emph{Mol. Phys.} \textbf{2013}, \emph{111}, 1271--1274\relax
\mciteBstWouldAddEndPuncttrue
\mciteSetBstMidEndSepPunct{\mcitedefaultmidpunct}
{\mcitedefaultendpunct}{\mcitedefaultseppunct}\relax
\EndOfBibitem
\bibitem[Hait and {Head-Gordon}(2021)Hait, and
  {Head-Gordon}]{haitOrbitalOptimizedDensity2021}
Hait,~D.; {Head-Gordon},~M. Orbital {{Optimized Density Functional Theory}} for
  {{Electronic Excited States}}. \emph{J. Phys. Chem. Lett.} \textbf{2021},
  \emph{12}, 4517--4529\relax
\mciteBstWouldAddEndPuncttrue
\mciteSetBstMidEndSepPunct{\mcitedefaultmidpunct}
{\mcitedefaultendpunct}{\mcitedefaultseppunct}\relax
\EndOfBibitem
\bibitem[Horbatenko \latin{et~al.}(2021)Horbatenko, Sadiq, Lee, Filatov, and
  Choi]{horbatenkoMixedReferenceSpinFlipTimeDependent2021}
Horbatenko,~Y.; Sadiq,~S.; Lee,~S.; Filatov,~M.; Choi,~C.~H. Mixed-{{Reference
  Spin-Flip Time-Dependent Density Functional Theory}} ({{MRSF-TDDFT}}) as a
  {{Simple}} yet {{Accurate Method}} for {{Diradicals}} and {{Diradicaloids}}.
  \emph{J. Chem. Theory Comput.} \textbf{2021}, \emph{17}, 848--859\relax
\mciteBstWouldAddEndPuncttrue
\mciteSetBstMidEndSepPunct{\mcitedefaultmidpunct}
{\mcitedefaultendpunct}{\mcitedefaultseppunct}\relax
\EndOfBibitem
\bibitem[Chen \latin{et~al.}(2017)Chen, Zhang, Jin, Yang, Su, and
  Yang]{chenMultireferenceDensityFunctional2017}
Chen,~Z.; Zhang,~D.; Jin,~Y.; Yang,~Y.; Su,~N.~Q.; Yang,~W. Multireference
  {{Density Functional Theory}} with {{Generalized Auxiliary Systems}} for
  {{Ground}} and {{Excited States}}. \emph{J. Phys. Chem. Lett.} \textbf{2017},
  \emph{8}, 4479--4485\relax
\mciteBstWouldAddEndPuncttrue
\mciteSetBstMidEndSepPunct{\mcitedefaultmidpunct}
{\mcitedefaultendpunct}{\mcitedefaultseppunct}\relax
\EndOfBibitem
\bibitem[{van Aggelen} \latin{et~al.}(2013){van Aggelen}, Yang, and
  Yang]{vanaggelenExchangecorrelationEnergyPairing2013}
{van Aggelen},~H.; Yang,~Y.; Yang,~W. Exchange-Correlation Energy from Pairing
  Matrix Fluctuation and the Particle-Particle Random-Phase Approximation.
  \emph{Phys. Rev. A} \textbf{2013}, \emph{88}, 030501\relax
\mciteBstWouldAddEndPuncttrue
\mciteSetBstMidEndSepPunct{\mcitedefaultmidpunct}
{\mcitedefaultendpunct}{\mcitedefaultseppunct}\relax
\EndOfBibitem
\bibitem[{van Aggelen} \latin{et~al.}(2014){van Aggelen}, Yang, and
  Yang]{vanaggelenExchangecorrelationEnergyPairing2014}
{van Aggelen},~H.; Yang,~Y.; Yang,~W. Exchange-Correlation Energy from Pairing
  Matrix Fluctuation and the Particle-Particle Random Phase Approximation.
  \emph{J. Chem. Phys.} \textbf{2014}, \emph{140}, 18A511\relax
\mciteBstWouldAddEndPuncttrue
\mciteSetBstMidEndSepPunct{\mcitedefaultmidpunct}
{\mcitedefaultendpunct}{\mcitedefaultseppunct}\relax
\EndOfBibitem
\bibitem[Yang \latin{et~al.}(2013)Yang, {van Aggelen}, Steinmann, Peng, and
  Yang]{yangBenchmarkTestsSpin2013}
Yang,~Y.; {van Aggelen},~H.; Steinmann,~S.~N.; Peng,~D.; Yang,~W. Benchmark
  Tests and Spin Adaptation for the Particle-Particle Random Phase
  Approximation. \emph{J. Chem. Phys.} \textbf{2013}, \emph{139}, 174110\relax
\mciteBstWouldAddEndPuncttrue
\mciteSetBstMidEndSepPunct{\mcitedefaultmidpunct}
{\mcitedefaultendpunct}{\mcitedefaultseppunct}\relax
\EndOfBibitem
\bibitem[Zhang and Yang(2016)Zhang, and
  Yang]{zhangAccurateEfficientCalculation2016}
Zhang,~D.; Yang,~W. Accurate and Efficient Calculation of Excitation Energies
  with the Active-Space Particle-Particle Random Phase Approximation. \emph{J.
  Chem. Phys.} \textbf{2016}, \emph{145}, 144105\relax
\mciteBstWouldAddEndPuncttrue
\mciteSetBstMidEndSepPunct{\mcitedefaultmidpunct}
{\mcitedefaultendpunct}{\mcitedefaultseppunct}\relax
\EndOfBibitem
\bibitem[Bannwarth \latin{et~al.}(2020)Bannwarth, Yu, Hohenstein, and
  Mart{\'i}nez]{bannwarthHoleHoleTamm2020}
Bannwarth,~C.; Yu,~J.~K.; Hohenstein,~E.~G.; Mart{\'i}nez,~T.~J.
  Hole\textendash Hole {{Tamm}}\textendash{{Dancoff-approximated}} Density
  Functional Theory: {{A}} Highly Efficient Electronic Structure Method
  Incorporating Dynamic and Static Correlation. \emph{J. Chem. Phys.}
  \textbf{2020}, \emph{153}, 024110\relax
\mciteBstWouldAddEndPuncttrue
\mciteSetBstMidEndSepPunct{\mcitedefaultmidpunct}
{\mcitedefaultendpunct}{\mcitedefaultseppunct}\relax
\EndOfBibitem
\bibitem[Tahir and Ren(2019)Tahir, and
  Ren]{tahirComparingParticleparticleParticlehole2019}
Tahir,~M.~N.; Ren,~X. Comparing Particle-Particle and Particle-Hole Channels of
  the Random Phase Approximation. \emph{Phys. Rev. B} \textbf{2019}, \emph{99},
  195149\relax
\mciteBstWouldAddEndPuncttrue
\mciteSetBstMidEndSepPunct{\mcitedefaultmidpunct}
{\mcitedefaultendpunct}{\mcitedefaultseppunct}\relax
\EndOfBibitem
\bibitem[Shao \latin{et~al.}(2003)Shao, {Head-Gordon}, and
  Krylov]{shaoSpinFlipApproach2003}
Shao,~Y.; {Head-Gordon},~M.; Krylov,~A.~I. The Spin\textendash Flip Approach
  within Time-Dependent Density Functional Theory: {{Theory}} and Applications
  to Diradicals. \emph{J. Chem. Phys.} \textbf{2003}, \emph{118},
  4807--4818\relax
\mciteBstWouldAddEndPuncttrue
\mciteSetBstMidEndSepPunct{\mcitedefaultmidpunct}
{\mcitedefaultendpunct}{\mcitedefaultseppunct}\relax
\EndOfBibitem
\bibitem[Wang and Ziegler(2004)Wang, and
  Ziegler]{wangTimedependentDensityFunctional2004}
Wang,~F.; Ziegler,~T. Time-Dependent Density Functional Theory Based on a
  Noncollinear Formulation of the Exchange-Correlation Potential. \emph{J.
  Chem. Phys.} \textbf{2004}, \emph{121}, 12191--12196\relax
\mciteBstWouldAddEndPuncttrue
\mciteSetBstMidEndSepPunct{\mcitedefaultmidpunct}
{\mcitedefaultendpunct}{\mcitedefaultseppunct}\relax
\EndOfBibitem
\bibitem[Bernard \latin{et~al.}(2012)Bernard, Shao, and
  Krylov]{bernardGeneralFormulationSpinflip2012}
Bernard,~Y.~A.; Shao,~Y.; Krylov,~A.~I. General Formulation of Spin-Flip
  Time-Dependent Density Functional Theory Using Non-Collinear Kernels:
  {{Theory}}, Implementation, and Benchmarks. \emph{J. Chem. Phys.}
  \textbf{2012}, \emph{136}, 204103\relax
\mciteBstWouldAddEndPuncttrue
\mciteSetBstMidEndSepPunct{\mcitedefaultmidpunct}
{\mcitedefaultendpunct}{\mcitedefaultseppunct}\relax
\EndOfBibitem
\bibitem[Li and Liu(2012)Li, and Liu]{liTheoreticalNumericalAssessments2012}
Li,~Z.; Liu,~W. Theoretical and Numerical Assessments of Spin-Flip
  Time-Dependent Density Functional Theory. \emph{J. Chem. Phys.}
  \textbf{2012}, \emph{136}, 024107\relax
\mciteBstWouldAddEndPuncttrue
\mciteSetBstMidEndSepPunct{\mcitedefaultmidpunct}
{\mcitedefaultendpunct}{\mcitedefaultseppunct}\relax
\EndOfBibitem
\bibitem[Jin \latin{et~al.}(2020)Jin, Su, Chen, and
  Yang]{jinIntroductoryLectureWhen2020}
Jin,~Y.; Su,~N.~Q.; Chen,~Z.; Yang,~W. Introductory Lecture: When the Density
  of the Noninteracting Reference System Is Not the Density of the Physical
  System in Density Functional Theory. \emph{Faraday Discuss.} \textbf{2020},
  \emph{224}, 9--26\relax
\mciteBstWouldAddEndPuncttrue
\mciteSetBstMidEndSepPunct{\mcitedefaultmidpunct}
{\mcitedefaultendpunct}{\mcitedefaultseppunct}\relax
\EndOfBibitem
\bibitem[Jin \latin{et~al.}(2017)Jin, Zhang, Chen, Su, and
  Yang]{jinGeneralizedOptimizedEffective2017}
Jin,~Y.; Zhang,~D.; Chen,~Z.; Su,~N.~Q.; Yang,~W. Generalized {{Optimized
  Effective Potential}} for {{Orbital Functionals}} and {{Self-Consistent
  Calculation}} of {{Random Phase Approximations}}. \emph{J. Phys. Chem. Lett.}
  \textbf{2017}, \emph{8}, 4746--4751\relax
\mciteBstWouldAddEndPuncttrue
\mciteSetBstMidEndSepPunct{\mcitedefaultmidpunct}
{\mcitedefaultendpunct}{\mcitedefaultseppunct}\relax
\EndOfBibitem
\bibitem[{Head-Gordon} and Pople(1988){Head-Gordon}, and
  Pople]{head-gordonOptimizationWaveFunction1988}
{Head-Gordon},~M.; Pople,~J.~A. Optimization of Wave Function and Geometry in
  the Finite Basis {{Hartree-Fock}} Method. \emph{J. Phys. Chem.}
  \textbf{1988}, \emph{92}, 3063--3069\relax
\mciteBstWouldAddEndPuncttrue
\mciteSetBstMidEndSepPunct{\mcitedefaultmidpunct}
{\mcitedefaultendpunct}{\mcitedefaultseppunct}\relax
\EndOfBibitem
\bibitem[Peverati and {Head-Gordon}(2013)Peverati, and
  {Head-Gordon}]{peveratiOrbitalOptimizedDoublehybrid2013}
Peverati,~R.; {Head-Gordon},~M. Orbital Optimized Double-Hybrid Density
  Functionals. \emph{J. Chem. Phys.} \textbf{2013}, \emph{139}, 024110\relax
\mciteBstWouldAddEndPuncttrue
\mciteSetBstMidEndSepPunct{\mcitedefaultmidpunct}
{\mcitedefaultendpunct}{\mcitedefaultseppunct}\relax
\EndOfBibitem
\bibitem[Yaffe and Goddard(1976)Yaffe, and
  Goddard]{yaffeOrbitalOptimizationElectronic1976}
Yaffe,~L.~G.; Goddard,~W.~A. Orbital Optimization in Electronic Wave Functions;
  Equations for Quadratic and Cubic Convergence of General Multiconfiguration
  Wave Functions. \emph{Phys. Rev. A} \textbf{1976}, \emph{13},
  1682--1691\relax
\mciteBstWouldAddEndPuncttrue
\mciteSetBstMidEndSepPunct{\mcitedefaultmidpunct}
{\mcitedefaultendpunct}{\mcitedefaultseppunct}\relax
\EndOfBibitem
\bibitem[Jin \latin{et~al.}(2019)Jin, Su, and
  Yang]{jinRenormalizedSinglesGreen2019}
Jin,~Y.; Su,~N.~Q.; Yang,~W. Renormalized {{Singles Green}}'s {{Function}} for
  {{Quasi-Particle Calculations}} beyond the {{G0W0 Approximation}}. \emph{J.
  Phys. Chem. Lett.} \textbf{2019}, \emph{10}, 447--452\relax
\mciteBstWouldAddEndPuncttrue
\mciteSetBstMidEndSepPunct{\mcitedefaultmidpunct}
{\mcitedefaultendpunct}{\mcitedefaultseppunct}\relax
\EndOfBibitem
\bibitem[Szabo and Ostlund(2012)Szabo, and
  Ostlund]{szaboModernQuantumChemistry2012}
Szabo,~A.; Ostlund,~N.~S. \emph{Modern {{Quantum Chemistry}}: {{Introduction}}
  to {{Advanced Electronic Structure Theory}}}; {Courier Corporation},
  2012\relax
\mciteBstWouldAddEndPuncttrue
\mciteSetBstMidEndSepPunct{\mcitedefaultmidpunct}
{\mcitedefaultendpunct}{\mcitedefaultseppunct}\relax
\EndOfBibitem
\bibitem[Slater(1930)]{slaterNoteHartreeMethod1930}
Slater,~J.~C. Note on {{Hartree}}'s {{Method}}. \emph{Phys. Rev.}
  \textbf{1930}, \emph{35}, 210--211\relax
\mciteBstWouldAddEndPuncttrue
\mciteSetBstMidEndSepPunct{\mcitedefaultmidpunct}
{\mcitedefaultendpunct}{\mcitedefaultseppunct}\relax
\EndOfBibitem
\bibitem[Li \latin{et~al.}(2021)Li, Chen, and
  Yang]{liRenormalizedSinglesGreen2021}
Li,~J.; Chen,~Z.; Yang,~W. Renormalized {{Singles Green}}'s {{Function}} in the
  {{T-Matrix Approximation}} for {{Accurate Quasiparticle Energy Calculation}}.
  \emph{J. Phys. Chem. Lett.} \textbf{2021}, \emph{12}, 6203--6210\relax
\mciteBstWouldAddEndPuncttrue
\mciteSetBstMidEndSepPunct{\mcitedefaultmidpunct}
{\mcitedefaultendpunct}{\mcitedefaultseppunct}\relax
\EndOfBibitem
\bibitem[Ren \latin{et~al.}(2011)Ren, Tkatchenko, Rinke, and
  Scheffler]{renRandomPhaseApproximationElectron2011}
Ren,~X.; Tkatchenko,~A.; Rinke,~P.; Scheffler,~M. Beyond the {{Random-Phase
  Approximation}} for the {{Electron Correlation Energy}}: {{The Importance}}
  of {{Single Excitations}}. \emph{Phys. Rev. Lett.} \textbf{2011}, \emph{106},
  153003\relax
\mciteBstWouldAddEndPuncttrue
\mciteSetBstMidEndSepPunct{\mcitedefaultmidpunct}
{\mcitedefaultendpunct}{\mcitedefaultseppunct}\relax
\EndOfBibitem
\bibitem[Yang \latin{et~al.}(2014)Yang, Peng, Lu, and
  Yang]{yangExcitationEnergiesParticleparticle2014}
Yang,~Y.; Peng,~D.; Lu,~J.; Yang,~W. Excitation Energies from Particle-Particle
  Random Phase Approximation: {{Davidson}} Algorithm and Benchmark Studies.
  \emph{J. Chem. Phys.} \textbf{2014}, \emph{141}, 124104\relax
\mciteBstWouldAddEndPuncttrue
\mciteSetBstMidEndSepPunct{\mcitedefaultmidpunct}
{\mcitedefaultendpunct}{\mcitedefaultseppunct}\relax
\EndOfBibitem
\bibitem[Yang \latin{et~al.}(2013)Yang, {van Aggelen}, and
  Yang]{yangDoubleRydbergCharge2013}
Yang,~Y.; {van Aggelen},~H.; Yang,~W. Double, {{Rydberg}} and Charge Transfer
  Excitations from Pairing Matrix Fluctuation and Particle-Particle Random
  Phase Approximation. \emph{J. Chem. Phys.} \textbf{2013}, \emph{139},
  224105\relax
\mciteBstWouldAddEndPuncttrue
\mciteSetBstMidEndSepPunct{\mcitedefaultmidpunct}
{\mcitedefaultendpunct}{\mcitedefaultseppunct}\relax
\EndOfBibitem
\bibitem[Yang and Wu(2002)Yang, and Wu]{yangDirectMethodOptimized2002}
Yang,~W.; Wu,~Q. Direct {{Method}} for {{Optimized Effective Potentials}} in
  {{Density-Functional Theory}}. \emph{Phys. Rev. Lett.} \textbf{2002},
  \emph{89}, 143002\relax
\mciteBstWouldAddEndPuncttrue
\mciteSetBstMidEndSepPunct{\mcitedefaultmidpunct}
{\mcitedefaultendpunct}{\mcitedefaultseppunct}\relax
\EndOfBibitem
\bibitem[Wu and Yang(2003)Wu, and Yang]{wuDirectOptimizationMethod2003}
Wu,~Q.; Yang,~W. A Direct Optimization Method for Calculating Density
  Functionals and Exchange\textendash Correlation Potentials from Electron
  Densities. \emph{J. Chem. Phys.} \textbf{2003}, \emph{118}, 2498--2509\relax
\mciteBstWouldAddEndPuncttrue
\mciteSetBstMidEndSepPunct{\mcitedefaultmidpunct}
{\mcitedefaultendpunct}{\mcitedefaultseppunct}\relax
\EndOfBibitem
\bibitem[Wu \latin{et~al.}(2005)Wu, Cohen, and
  Yang]{wuAnalyticEnergyGradients2005}
Wu,~Q.; Cohen,~A.~J.; Yang,~W. Analytic Energy Gradients of the Optimized
  Effective Potential Method. \emph{J. Chem. Phys.} \textbf{2005}, \emph{123},
  134111\relax
\mciteBstWouldAddEndPuncttrue
\mciteSetBstMidEndSepPunct{\mcitedefaultmidpunct}
{\mcitedefaultendpunct}{\mcitedefaultseppunct}\relax
\EndOfBibitem
\bibitem[Zhang \latin{et~al.}(2014)Zhang, Peng, Zhang, and
  Yang]{zhangAnalyticGradientsGeometry2014}
Zhang,~D.; Peng,~D.; Zhang,~P.; Yang,~W. Analytic Gradients, Geometry
  Optimization and Excited State Potential Energy Surfaces from the
  Particle-Particle Random Phase Approximation. \emph{Phys. Chem. Chem. Phys.}
  \textbf{2014}, \emph{17}, 1025--1038\relax
\mciteBstWouldAddEndPuncttrue
\mciteSetBstMidEndSepPunct{\mcitedefaultmidpunct}
{\mcitedefaultendpunct}{\mcitedefaultseppunct}\relax
\EndOfBibitem
\bibitem[Martin \latin{et~al.}(2016)Martin, Reining, and
  Ceperley]{martinInteractingElectrons2016}
Martin,~R.~M.; Reining,~L.; Ceperley,~D.~M. \emph{Interacting {{Electrons}}};
  {Cambridge University Press}, 2016\relax
\mciteBstWouldAddEndPuncttrue
\mciteSetBstMidEndSepPunct{\mcitedefaultmidpunct}
{\mcitedefaultendpunct}{\mcitedefaultseppunct}\relax
\EndOfBibitem
\bibitem[Davidson(1975)]{davidsonIterativeCalculationFew1975}
Davidson,~E.~R. The Iterative Calculation of a Few of the Lowest Eigenvalues
  and Corresponding Eigenvectors of Large Real-Symmetric Matrices. \emph{J.
  Comput. Phys} \textbf{1975}, \emph{17}, 87--94\relax
\mciteBstWouldAddEndPuncttrue
\mciteSetBstMidEndSepPunct{\mcitedefaultmidpunct}
{\mcitedefaultendpunct}{\mcitedefaultseppunct}\relax
\EndOfBibitem
\bibitem[Stratmann \latin{et~al.}(1998)Stratmann, Scuseria, and
  Frisch]{stratmannEfficientImplementationTimedependent1998}
Stratmann,~R.~E.; Scuseria,~G.~E.; Frisch,~M.~J. An Efficient Implementation of
  Time-Dependent Density-Functional Theory for the Calculation of Excitation
  Energies of Large Molecules. \emph{J. Chem. Phys.} \textbf{1998}, \emph{109},
  8218--8224\relax
\mciteBstWouldAddEndPuncttrue
\mciteSetBstMidEndSepPunct{\mcitedefaultmidpunct}
{\mcitedefaultendpunct}{\mcitedefaultseppunct}\relax
\EndOfBibitem
\bibitem[Zhao \latin{et~al.}(1994)Zhao, Morrison, and
  Parr]{zhaoElectronDensitiesKohnSham1994}
Zhao,~Q.; Morrison,~R.~C.; Parr,~R.~G. From Electron Densities to {{Kohn-Sham}}
  Kinetic Energies, Orbital Energies, Exchange-Correlation Potentials, and
  Exchange-Correlation Energies. \emph{Phys. Rev. A} \textbf{1994}, \emph{50},
  2138--2142\relax
\mciteBstWouldAddEndPuncttrue
\mciteSetBstMidEndSepPunct{\mcitedefaultmidpunct}
{\mcitedefaultendpunct}{\mcitedefaultseppunct}\relax
\EndOfBibitem
\bibitem[Press \latin{et~al.}(1992)Press, Teukolsky, Vetterling, and
  Flannery]{pressNumericalRecipes2nd1992}
Press,~W.~H.; Teukolsky,~S.~A.; Vetterling,~W.~T.; Flannery,~B.~P.
  \emph{Numerical Recipes in {{C}} (2nd Ed.): The Art of Scientific Computing};
  {Cambridge University Press}: {USA}, 1992\relax
\mciteBstWouldAddEndPuncttrue
\mciteSetBstMidEndSepPunct{\mcitedefaultmidpunct}
{\mcitedefaultendpunct}{\mcitedefaultseppunct}\relax
\EndOfBibitem
\bibitem[Yang \latin{et~al.}(2004)Yang, Ayers, and
  Wu]{yangPotentialFunctionalsDual2004}
Yang,~W.; Ayers,~P.~W.; Wu,~Q. Potential {{Functionals}}: {{Dual}} to {{Density
  Functionals}} and {{Solution}} to the $v$-{{Representability Problem}}.
  \emph{Phys. Rev. Lett.} \textbf{2004}, \emph{92}, 146404\relax
\mciteBstWouldAddEndPuncttrue
\mciteSetBstMidEndSepPunct{\mcitedefaultmidpunct}
{\mcitedefaultendpunct}{\mcitedefaultseppunct}\relax
\EndOfBibitem
\bibitem[qm4()]{qm4d}
See http://www.qm4d.info for an in-house program for QM/MM simulations\relax
\mciteBstWouldAddEndPuncttrue
\mciteSetBstMidEndSepPunct{\mcitedefaultmidpunct}
{\mcitedefaultendpunct}{\mcitedefaultseppunct}\relax
\EndOfBibitem
\bibitem[Dunning(1989)]{dunningGaussianBasisSets1989}
Dunning,~T.~H. Gaussian Basis Sets for Use in Correlated Molecular
  Calculations. {{I}}. {{The}} Atoms Boron through Neon and Hydrogen. \emph{J.
  Chem. Phys.} \textbf{1989}, \emph{90}, 1007--1023\relax
\mciteBstWouldAddEndPuncttrue
\mciteSetBstMidEndSepPunct{\mcitedefaultmidpunct}
{\mcitedefaultendpunct}{\mcitedefaultseppunct}\relax
\EndOfBibitem
\bibitem[Li and Evangelista(2017)Li, and
  Evangelista]{liDrivenSimilarityRenormalization2017}
Li,~C.; Evangelista,~F.~A. Driven Similarity Renormalization Group:
  {{Third-order}} Multireference Perturbation Theory. \emph{J. Chem. Phys.}
  \textbf{2017}, \emph{146}, 124132\relax
\mciteBstWouldAddEndPuncttrue
\mciteSetBstMidEndSepPunct{\mcitedefaultmidpunct}
{\mcitedefaultendpunct}{\mcitedefaultseppunct}\relax
\EndOfBibitem
\bibitem[Li and Paldus(2003)Li, and
  Paldus]{liAccurateDeterminationRovibrational2003}
Li,~X.; Paldus,~J. An Accurate Determination of Rovibrational Spectra Using the
  Externally Corrected Coupled-Cluster Approaches: {{LiH}} Ground State.
  \emph{J. Chem. Phys.} \textbf{2003}, \emph{118}, 2470--2481\relax
\mciteBstWouldAddEndPuncttrue
\mciteSetBstMidEndSepPunct{\mcitedefaultmidpunct}
{\mcitedefaultendpunct}{\mcitedefaultseppunct}\relax
\EndOfBibitem
\bibitem[Abrams and Sherrill(2003)Abrams, and
  Sherrill]{abramsAssessmentAccuracyMultireference2003}
Abrams,~M.~L.; Sherrill,~C.~D. An {{Assessment}} of the {{Accuracy}} of
  {{Multireference Configuration Interaction}} ({{MRCI}}) and
  {{Complete-Active-Space Second-Order Perturbation Theory}} ({{CASPT2}}) for
  {{Breaking Bonds}} to {{Hydrogen}}. \emph{J. Phys. Chem. A} \textbf{2003},
  \emph{107}, 5611--5616\relax
\mciteBstWouldAddEndPuncttrue
\mciteSetBstMidEndSepPunct{\mcitedefaultmidpunct}
{\mcitedefaultendpunct}{\mcitedefaultseppunct}\relax
\EndOfBibitem
\bibitem[Biglari \latin{et~al.}(2014)Biglari, Shayesteh, and
  Maghari]{biglariInitioPotentialEnergy2014}
Biglari,~Z.; Shayesteh,~A.; Maghari,~A. Ab Initio Potential Energy Curves and
  Transition Dipole Moments for the Low-Lying States of {{CH}}+. \emph{Comput.
  Theor. Chem.} \textbf{2014}, \emph{1047}, 22--29\relax
\mciteBstWouldAddEndPuncttrue
\mciteSetBstMidEndSepPunct{\mcitedefaultmidpunct}
{\mcitedefaultendpunct}{\mcitedefaultseppunct}\relax
\EndOfBibitem
\bibitem[Jiang \latin{et~al.}(2012)Jiang, Jeffrey, and
  Wilson]{jiangEmpiricalCorrectionNondynamical2012}
Jiang,~W.; Jeffrey,~C.~C.; Wilson,~A.~K. Empirical {{Correction}} of
  {{Nondynamical Correlation Energy}} for {{Density Functionals}}. \emph{J.
  Phys. Chem. A} \textbf{2012}, \emph{116}, 9969--9978\relax
\mciteBstWouldAddEndPuncttrue
\mciteSetBstMidEndSepPunct{\mcitedefaultmidpunct}
{\mcitedefaultendpunct}{\mcitedefaultseppunct}\relax
\EndOfBibitem
\bibitem[Golze \latin{et~al.}(2020)Golze, Keller, and
  Rinke]{golzeAccurateAbsoluteRelative2020}
Golze,~D.; Keller,~L.; Rinke,~P. Accurate {{Absolute}} and {{Relative
  Core-Level Binding Energies}} from {{GW}}. \emph{J. Phys. Chem. Lett.}
  \textbf{2020}, \emph{11}, 1840--1847\relax
\mciteBstWouldAddEndPuncttrue
\mciteSetBstMidEndSepPunct{\mcitedefaultmidpunct}
{\mcitedefaultendpunct}{\mcitedefaultseppunct}\relax
\EndOfBibitem
\bibitem[Frisch \latin{et~al.}(2016)Frisch, Trucks, Schlegel, Scuseria, Robb,
  Cheeseman, Scalmani, Barone, Petersson, Nakatsuji, Li, Caricato, Marenich,
  Bloino, Janesko, Gomperts, Mennucci, Hratchian, Ortiz, Izmaylov, Sonnenberg,
  Williams-Young, Ding, Lipparini, Egidi, Goings, Peng, Petrone, Henderson,
  Ranasinghe, Zakrzewski, Gao, Rega, Zheng, Liang, Hada, Ehara, Toyota, Fukuda,
  Hasegawa, Ishida, Nakajima, Honda, Kitao, Nakai, Vreven, Throssell,
  Montgomery, Peralta, Ogliaro, Bearpark, Heyd, Brothers, Kudin, Staroverov,
  Keith, Kobayashi, Normand, Raghavachari, Rendell, Burant, Iyengar, Tomasi,
  Cossi, Millam, Klene, Adamo, Cammi, Ochterski, Martin, Morokuma, Farkas,
  Foresman, and Fox]{g16}
Frisch,~M.~J.; Trucks,~G.~W.; Schlegel,~H.~B.; Scuseria,~G.~E.; Robb,~M.~A.;
  Cheeseman,~J.~R.; Scalmani,~G.; Barone,~V.; Petersson,~G.~A.; Nakatsuji,~H.
  \latin{et~al.}  Gaussian16 {R}evision A.03. 2016; Gaussian Inc. Wallingford
  CT\relax
\mciteBstWouldAddEndPuncttrue
\mciteSetBstMidEndSepPunct{\mcitedefaultmidpunct}
{\mcitedefaultendpunct}{\mcitedefaultseppunct}\relax
\EndOfBibitem
\bibitem[Weigend(2006)]{weigendAccurateCoulombfittingBasis2006}
Weigend,~F. Accurate {{Coulomb-fitting}} Basis Sets for {{H}} to {{Rn}}.
  \emph{Phys. Chem. Chem. Phys.} \textbf{2006}, \emph{8}, 1057\relax
\mciteBstWouldAddEndPuncttrue
\mciteSetBstMidEndSepPunct{\mcitedefaultmidpunct}
{\mcitedefaultendpunct}{\mcitedefaultseppunct}\relax
\EndOfBibitem
\bibitem[Ren \latin{et~al.}(2012)Ren, Rinke, Blum, Wieferink, Tkatchenko,
  Sanfilippo, Reuter, and
  Scheffler]{renResolutionofidentityApproachHartree2012}
Ren,~X.; Rinke,~P.; Blum,~V.; Wieferink,~J.; Tkatchenko,~A.; Sanfilippo,~A.;
  Reuter,~K.; Scheffler,~M. Resolution-of-Identity Approach to
  {{Hartree}}\textendash{{Fock}}, Hybrid Density Functionals, {{RPA}}, {{MP2
  and GW with}} Numeric Atom-Centered Orbital Basis Functions. \emph{New J.
  Phys.} \textbf{2012}, \emph{14}, 053020\relax
\mciteBstWouldAddEndPuncttrue
\mciteSetBstMidEndSepPunct{\mcitedefaultmidpunct}
{\mcitedefaultendpunct}{\mcitedefaultseppunct}\relax
\EndOfBibitem
\bibitem[Eichkorn \latin{et~al.}(1995)Eichkorn, Treutler, {\"O}hm, H{\"a}ser,
  and Ahlrichs]{eichkornAuxiliaryBasisSets1995}
Eichkorn,~K.; Treutler,~O.; {\"O}hm,~H.; H{\"a}ser,~M.; Ahlrichs,~R. Auxiliary
  Basis Sets to Approximate {{Coulomb}} Potentials. \emph{Chem. Phys. Lett.}
  \textbf{1995}, \emph{240}, 283--290\relax
\mciteBstWouldAddEndPuncttrue
\mciteSetBstMidEndSepPunct{\mcitedefaultmidpunct}
{\mcitedefaultendpunct}{\mcitedefaultseppunct}\relax
\EndOfBibitem
\bibitem[Feller(1996)]{fellerRoleDatabasesSupport1996}
Feller,~D. The Role of Databases in Support of Computational Chemistry
  Calculations. \emph{J. Comput. Chem.} \textbf{1996}, \emph{17},
  1571--1586\relax
\mciteBstWouldAddEndPuncttrue
\mciteSetBstMidEndSepPunct{\mcitedefaultmidpunct}
{\mcitedefaultendpunct}{\mcitedefaultseppunct}\relax
\EndOfBibitem
\bibitem[Schuchardt \latin{et~al.}(2007)Schuchardt, Didier, Elsethagen, Sun,
  Gurumoorthi, Chase, Li, and Windus]{schuchardtBasisSetExchange2007}
Schuchardt,~K.~L.; Didier,~B.~T.; Elsethagen,~T.; Sun,~L.; Gurumoorthi,~V.;
  Chase,~J.; Li,~J.; Windus,~T.~L. Basis {{Set Exchange}}:\, {{A Community
  Database}} for {{Computational Sciences}}. \emph{J. Chem. Inf. Model.}
  \textbf{2007}, \emph{47}, 1045--1052\relax
\mciteBstWouldAddEndPuncttrue
\mciteSetBstMidEndSepPunct{\mcitedefaultmidpunct}
{\mcitedefaultendpunct}{\mcitedefaultseppunct}\relax
\EndOfBibitem
\end{mcitethebibliography}

\end{document}